\documentclass[journal=jctcce]{achemso}

\usepackage[english]{babel}
\usepackage{amsmath,bm}
\usepackage{amssymb}
\usepackage{mathrsfs}
\usepackage{amsfonts}
\usepackage{graphicx}
\usepackage{siunitx}
\usepackage{xfrac}
\usepackage[version=3]{mhchem}
\usepackage[acronym]{glossaries}

\DeclareSIUnit\au{\text {au}}

\newcommand{\ev}[1]{\langle #1 \rangle}

\newacronym{cboa}{CBOA}{cavity Born-Oppenheimer approximation}
\newacronym{bo}{BOA}{Born-Oppenheimer approximation}
\newacronym{cbohf}{CBO-HF}{cavity Born-Oppenheimer Hartree-Fock}
\newacronym{scf}{SCF}{self-consistent field}
\newacronym{dse}{DSE}{dipole self-energy}
\newacronym{pes}{PES}{potential energy surface}
\newacronym{cpes}{cPES}{cavity potential energy surface}
\newacronym{vsc}{VSC}{vibrational-strong coupling}
\newacronym{esc}{ESC}{electronic-strong coupling}
\newacronym{lp}{LP}{lower polariton}
\newacronym{up}{UP}{upper polariton}
\newacronym{cphf}{CPHF}{coupled-perturbed Hartree-Fock}

\title{Molecular Polarizability under Vibrational Strong Coupling}

\author{Thomas Schnappinger}
\email{thomas.schnappinger@fysik.su.se}
\affiliation{Department of Physics, Stockholm University, AlbaNova University Center, SE-106 91 Stockholm, Sweden}
\author{Markus Kowalewski}
\email{markus.kowalewski@fysik.su.se}
\affiliation{Department of Physics, Stockholm University, AlbaNova University Center, SE-106 91 Stockholm, Sweden}

\begin{document}

\begin{abstract}
Polaritonic chemistry offers the possibility of modifying molecular properties and even influencing chemical reactivity through strong coupling between vibrational transitions and confined light modes in optical cavities.
Despite considerable theoretical progress and due to the complexity of the coupled light-matter system, the fundamental mechanism how and if collective strong coupling can induce local changes of individual molecules is still unclear. 
We derive an analytical formulation of static polarizabilities within linear response theory for molecules under strong coupling using the cavity Born-Oppenheimer Hartree-Fock ansatz.
This ab-initio method consistently describes vibrational strong coupling and electron-photon interactions even for ensembles of molecules. 
For different types of molecular ensemble, we observed local changes in the polarizabilities and dipole moments that are induced by collective strong coupling.
Furthermore, we used the polarizabilities to calculate vibro-polaritonic Raman spectra in the harmonic approximation.
This allows us to comprehensively compare the effect of vibrational strong coupling on IR and Raman spectra on an equal footing.
\end{abstract}

\maketitle
\section{Introduction}

By placing molecules in a modified photonic environment, such as a Fabry-P\'erot cavity, hybrid states known as polaritons can be formed.
These polaritons are hybrid light–matter states that emerge from the resonant exchange of energy between an optically bright transition in matter and the confined photonic mode of the optical cavity~\cite{Fregoni2022-op,Dunkelberger2022-oh,Ebbesen2023-fd,Bhuyan2023-se}.
The spectroscopic signature of such a strong light-matter coupling is the formation of a \gls{lp} transition and a \gls{up} transition separated by the Rabi splitting frequency, $\Omega_{R}$, which result from a distinct splitting of the peak in the absorption spectrum~\cite{Dovzhenko2018-ph,Hertzog2019-uh}.
A particularly fascinating phenomenon occurs when the light field is strongly coupled to molecular vibrations, called \gls{vsc}, where it seems possible to influence ground state reactions~\cite{Thomas2016-fy,Lather2019-yn,Hirai2020-uv,Ahn2023-qk,Patrahau2024-zx,Brawley2025-qh} in the "dark", i.e. without external illumination.
Despite considerable experimental and theoretical efforts in recent years~\cite{Schutz2020-en,Sidler2022-cg,Taylor2022-pv,Sanchez-Barquilla2022-dq,Simpkins2023-ze,Schnappinger2023-hh,Svendsen2023-ly,Sidler2024-tw,Michon2024-vb,Nelson2024-ym,Ying2024-aj,Sanchez_Martinez2024-wu,Vasil2025-by}, 
a fundamental understanding of the underlying microscopic and macroscopic physical mechanisms of \gls{vsc} is still lacking.
An atomistic understanding of the phenomenon is challenging due to the complexity of the collective nature.
However, a self-consistent treatment of the field-dressed electronic structure in a molecular ensemble appears to be a crucial ingredient in the description of cavity-induced microscopic changes under collective strong coupling~\cite{Haugland2020-xh,Schnappinger2023-hh,Angelico2023-jh,Schnappinger2023-wp,Sidler2024-tw,Fischer2024-jb,Castagnola2024-vt}.
The \gls{cbohf} approach~\cite{Schnappinger2023-hh} is a formulation of the well-known Harte-Fock ansatz in the \gls{cboa}~\cite{flick2017atoms,flick2017cavity,Flick2018-ns} that treats the dressed electron-photon system in an ab-initio fashion.
Using the \gls{cbohf} framework, we have been able to identify local strong coupling effects in molecular ensembles~\cite{Schnappinger2023-hh,Sidler2024-tw} and simulate vibro-polaritonic IR spectra within the harmonic approximation~\cite{Schnappinger2023-wp}.

The molecular polarizability $\bm{\alpha}$ is an important property that describes how the electronic structure reacts under the influence of an external electric field.
For example, the change in polarizability determines whether a vibrational mode is Raman active or not~\cite{Bernath2005-bt}.
In addition to defining light-matter interactions, molecular polarizabilities also influence how molecules interact by defining dispersion interactions~\cite{London1937-um,Maitland1983IntermolecularFT}, which are, for example, essential for understanding solvation processes.
The maxima of the static polarizability along a bond dissociation coordinate can be used to define cut-off points for bond cleavage~\cite{Hait2023-do}, as they represent the beginning of the localization of the shared electron density in the constituent fragments.
Furthermore, polarizabilities are relevant in the context of strong light-matter coupling.
It has been experimentally demonstrated~\cite{Patrahau2024-zx} that \gls{vsc} can modify the London dispersion forces, or, in other words, the underlying molecular polarizability under collective strong coupling. 
Using static polarizabilities, it is possible to estimate the influence of cavity interactions on molecular geometries~\cite{Schnappinger2024-vt}.

Within the framework of the \gls{cbohf} ansatz, we derive an analytic expression for the static polarizability $\bm{\alpha}$ as the second derivative of the energy with respect to an external electric field using linear-response theory. 
To validate the linear-response approach, we compare $\bm{\alpha}$ with numerically computed values by explicitly including an external electric field in the electronic \gls{cboa} Hamiltonian and solving the \gls{cbohf} equation.
We then investigate how polarizabilities and permanent dipole moments change under \gls{vsc} in the case of a single molecular system and small molecular ensembles for a set of molecules. 
In the first part, we investigate whether cavity-mediated collective interactions modify the dipole moments and polarizabilities of individual molecules in an ensemble.
This type of local modification of $\bm{\alpha}$ under \gls{vsc} has recently been shown for a full-harmonic model~\cite{Horak2024-fa}.  
In the second part of the manuscript, we use the analytical polarizability to extend our recent work on ab-initio vibro-polaritonic spectra~\cite{Schnappinger2023-wp} to determine Raman spectra for molecules under \gls{vsc}.
Raman scattering is a useful method for obtaining information about material properties and chemical structures, especially when probing the rovibrational structure.
However, the influence of \gls{vsc} on Raman scattering is still under discussion on both the experimental~\cite{Shalabney2015-qs,Takele2021-xv,Verdelli2022-yl} and theoretical~\cite{del-Pino2015-un,Strashko2016-nt,Huang2024-is,Huang2025-lu} side.
Here, we want to investigate the different effects of \gls{vsc} on IR and Raman spectra for a single formaldehyde molecule coupled to two cavity modes in a simplified Fabry-P\'erot-like setup.

\section{Theory}

The starting point for ab-initio studies of \gls{vsc} is the \gls{cboa}~\cite{flick2017atoms,flick2017cavity,Flick2018-ns} in the length gauge and dipole approximation.
Within \gls{cboa} the cavity field modes are grouped with the nuclei in a generalized Born-Huang expansion~\cite{Schafer2018-vf,Ruggenthaler2023-aa} which allows to first solve the quantum problem of the electrons for a fixed nuclear and photonic configuration.
The combined nuclear-photonic problem can subsequently be solved fully quantum mechanically or semi-classically.
The electronic \gls{cboa} Hamiltonian for $N_{M}$ cavity field modes after the adiabatic separation reads 
\begin{equation}
 \label{eq:HCBO}
\hat{H}_{\mathrm{CBO}} = \hat{H}_{el} + \sum_{c=1}^{N_{M}} \frac{\omega_{c}^2}{2} \left(\hat{q}_{c} -\frac{\hat{d}_c }{\omega_{c}}\right)^2
\end{equation}
where $\hat{H}_{el}$ is the Hamiltonian for the field-free many-electron system, $q_{c}$ is the photon displacement coordinate, $\omega_c$ is the frequency of the cavity mode $c$, and $N_{M}$ is the number of modes.
The linear coupling between the molecular system and the photon displacement field, as well as the \gls{dse} contribution~\cite{Rokaj2018-ww,Schafer2020-cb}, which describes the self-polarization of the molecule-cavity system, is characterized by the dressed dipole operator $\hat{d}_c$. This operator is defined as scaler product of the standard dipole operator $\bm{\hat{\mu}}$ and the coupling parameter $\bm{\lambda}_{c}$:
 \begin{equation}
 \label{eq:lam}
\hat{d}_c = \boldsymbol{\lambda}_c \cdot  \bm{\hat{\mu}} =  \boldsymbol{\lambda}_c \cdot \left(  \bm{\hat{\mu}}_{el} +  \bm{\mu}_{nuc}\right) \quad  \text{ with} \quad 
 \bm{\lambda}_{c} =  \bm{e}_{c} \lambda_{c} =  \bm{e}_{c}  \sqrt{\frac{4 \pi}{\mathcal{V}_{c}}}\,.
\end{equation}
The unit vector $\bm{e}_{c}$ denotes the polarization axis of the cavity mode and $\lambda_{c}$ is defined by the effective mode volume $\mathcal{V}_{c}$ of the corresponding cavity mode. 
For simplicity, we assume that $\mathcal{V}$ corresponds to the physical volume $V$.
To go beyond the case of a single molecule, the collective coupling strength $\bm{\lambda}_{c}$ is kept constant by applying a scaling factor of ${1}/{\sqrt{N_{mol}}}$ to obtain a fixed Rabi splitting for different ensemble sizes:
\begin{equation}
\label{eq:coupling}
\bm{\lambda}_{c} = \bm{e}_c \frac{\lambda_0}{\sqrt{N_{mol}}} 
\end{equation} 
Here, $\lambda_0$ is treated as a tunable coupling parameter and is equivalent to $\lambda_c$ in Eq.~\eqref{eq:lam} for a single molecule. 
As a result of the rescaling, we increase the physical volume $V$ of the cavity, but by including more molecules, we effectively keep the average density of molecules $N_{mol}/V$ fixed.

In our recent work~\cite{Schnappinger2023-hh,Schnappinger2023-wp} we have introduced the \gls{cbohf} approach, which represents a formulation of the well-known Hartree-Fock ansatz in the context of the \gls{cboa}. As in standard Hartree-Fock the many-electron wave function is a single Slater determinant $\Psi$ which is optimized for a fixed nuclear configuration and a set of photon displacement coordinates.
In the following we reformulate the optimization in terms of an exponential parametrization:
\begin{equation}
    \big| \Psi (\bm{\kappa}) \bigr\rangle = e^{\hat{\kappa}} \big| \Psi_0 \bigr\rangle
\end{equation}
where $e^{\hat{\kappa}}$ is a unitary operator that performs rotations between occupied and virtual spin orbitals in a reference determinant wavefunction $\Psi_0$.
The single excitation operator $\hat{\kappa}$ in the second-quantization formalism can be written as
\begin{equation}
    \hat{\kappa} = \sum_a^{occ}\sum_r^{virt} \left( \kappa_{ar} \hat{a}^{\dagger}_a\hat{a}_r - \kappa_{ar} \hat{a}^{\dagger}_r\hat{a}_a \right)
\end{equation}
where $\hat{a}^{\dagger}_k$ and $\hat{a}_k$ are the fermionic creation and annihilation operators of the spin orbital $k$, respectively, and $\kappa_{ar}$ are the orbital rotation parameters.
In each optimization step, the orbitals in the reference determinant $\Psi_0$ are updated so that variations of the orbital rotation parameters around $\bm{\kappa}=0$ are always considered.
The converged \gls{cbohf} energy then reads:
\begin{equation}
    E_{\mathrm{CBO}} \left(\bm{\kappa}\right) = \bigl\langle \Psi \left( \bm{\kappa} \right) \big| \hat{H}_{\mathrm{CBO}} \big| \Psi \left( \bm{\kappa} \right) \bigr\rangle 
\end{equation}
Using the converged Slater determinant $\Psi \left( \bm{\kappa} \right)$,
the expectation values of the dipole moment~\cite{aszabo82-qc} can be
calculated with the corresponding dipole operators
\begin{equation}
    \langle\bm{\hat{\mu}}\rangle_{\mathrm{CBO}} = \langle \Psi \left( \bm{\kappa} \right) \big| \bm{\hat{\mu}}_{el}\big| \Psi \left( \bm{\kappa} \right) \rangle + \bm{\mu}_{nuc} = - \sum_a^{occ} \langle a |\bm{\hat{r}} | a \rangle + \sum_A^{N_{nuc}} Z_{A}\bm{R}_A
\end{equation}
where $a$ are the occupied spin orbitals, $\bm{\hat{r}}$ and $\bm{R}_A$ are the electronic and nuclear position vectors, and $Z_{A}$ is the nuclear charge. 

\subsection{\label{subsec:alpha}\gls{cbohf} polarizability}

The definition of the static polarizability tensor $\bm{\alpha}$ of a molecular system is the response of its dipole moment $\langle\bm{\hat{\mu}}\rangle$ to a static external electric field $\bm{F}$, which is equivalent to the second derivative of the electronic energy $E$ with respect to $\bm{F}$~\cite{atkins_molecular_2011}:
\begin{equation}
    \ev{\alpha_{ij}} = \frac{\partial \langle\hat{\mu}_{i}\rangle}{\partial F_{j}} = \frac{\partial^2 E}{\partial F_{i} \partial F_{j}}
\end{equation}
This polarizability can be calculated self-consistently by explicitly including $\bm{F}$ in the electronic \gls{cboa} Hamiltonian:
\begin{equation}
 \label{eq:HCBO_ext}
\hat{H}_{\mathrm{CBO}}^{\bm{F}} = \hat{H}_{el} + \sum_{c=1}^{N_{M}} \frac{\omega_{c}^2}{2} \left(\hat{q}_{c} -\frac{\hat{d}_c}{\omega_{c}}\right)^2 - \left( \bm{\hat{\mu}} + \bm{\mu}_{nuc}\right)\bm{F}
\end{equation}
The converged \gls{cbohf} energy then reads:
\begin{equation}
    E_{\mathrm{CBO}}^{\bm{F}} \left(\bm{\kappa}\right) = \bigl\langle \Phi \left( \bm{\kappa} \right) \big| \hat{H}_{\mathrm{CBO}}^{\mathrm{ext}} \big| \Phi \left( \bm{\kappa} \right) \bigr\rangle 
\end{equation}
The converged Slater determinant $\Phi \left( \bm{\kappa} \right)$ explicitly taking into account  $\bm{F}$ is then used to determine $\ev{\bm{\hat{\mu}}}_{\mathrm{CBO}}^{\bm{F}} = \langle \Phi \left( \bm{\kappa}\right) \vert \bm{\hat{\mu}} \vert \Phi \left( \bm{\kappa}\right)\rangle$. 
The \gls{cbohf} polarizability $\ev{\bm{\alpha}}_{\mathrm{CBO}}^{\mathrm{num}}$ can be approximated numerically via finite differences:
\begin{equation}
\ev{\alpha_{ij}}_{\mathrm{CBO}}^{\mathrm{num}} \approx \frac{\ev{\hat{\mu}_{i}}_{\mathrm{CBO}}^{\mathrm{F_{j}}}\  - \ \ev{\hat{\mu}_{i}}_{\mathrm{CBO}}^{\mathrm{-F_{j}}} }{ 2F_{j}}
\end{equation}
It should be noted that this formula has a leading error term that is quadratic in the applied field and depending on the used field strength the results can be contaminated by higher order responses (hyperpolarizabilities)~\cite{Brakestad2020-bo}.

Alternatively, the polarizability $\ev{\bm{\alpha}}_{\mathrm{CBO}}^{\mathrm{cphf}}$ can be calculated using linear-response theory~\cite{Helgaker1988-if}. 
\begin{equation}
\label{eq:a_cphf}
\ev{\alpha_{ij}}_{\mathrm{CBO}}^{\mathrm{cphf}} = - \left[ \frac{\partial^2 E_{\mathrm{CBO}}^{0}}{\partial F_{i}\partial F_{j}} +\sum_{a}^{occ}\sum_{r}^{virt} \frac{\partial^2 E_{\mathrm{CBO}}^{0}}{\partial F_{i}\partial{\kappa}_{ar}}\Bigg|_{\kappa=0} \frac{\partial\kappa_{ar}}{\partial F_{j}}\right]
\end{equation}
Here $E_{\mathrm{CBO}}^{0} = \bigl\langle \Psi \left( \bm{\kappa} \right) \big| \hat{H}_{\mathrm{cbo}}^{\mathrm{ext}} \big| \Psi \left( \bm{\kappa} \right) \bigr\rangle$ is the \gls{cbohf} expectation value of $\hat{H}_{\mathrm{cbo}}^{\mathrm{ext}}$ calculated using the Slater determinant $\Psi \left( \bm{\kappa}\right)$ optimized without the external field. 
Since the explicit dependence of $E_{\mathrm{CBO}}^{0}$ on the electric field is linear (see the first term in Eq.~\ref{eq:a_cphf}), its second derivative with respect to $\bm{F}$ is zero. 
The second term interpretable as perturbed electronic gradient is
\begin{equation}
\frac{\partial^2 E_{\mathrm{CBO}}^{0}}{\partial F_{i}\partial\kappa_{ar}}\Bigg|_{\bm{\kappa}=0} = 2\frac{\bigl\langle \Psi_a^r \big|\hat{H}_{\mathrm{CBO}}^{\mathrm{ext}} \big| \Psi \left( \bm{\kappa} \right) \bigr\rangle}{\partial F_{i}} = -2\bigl\langle r \big| \hat{\mu}_i \big| a \bigr\rangle \,.
\end{equation}
where $a$ and $r$ are an occupied spin orbital and a virtual spin orbital, $\Psi \left( \bm{\kappa} \right)$ is the \gls{cbohf} Slater determinant optimized without the external field and $\Psi_a^r$ is the corresponding single excited determinant.
Due to Brillouin's theorem, all contributions of $\hat{H}_{\mathrm{CBO}}^{\mathrm{ext}}$ except the dipole interaction with the external field are zero.
The only remaining part is the wave-function linear-response vector $\partial\kappa_{ar}/ F_{j}$, which requires solving the linear-response equations that are the \gls{cbohf} counterparts of the \gls{cphf} equations in the cavity-free case. These \gls{cbohf} linear-response equations reads:
\begin{equation}
\sum_b^{occ}\sum_s^{virt} \left(A_{ar,bs} + B_{ar,bs}\right) \frac{\partial\kappa_{ar}}{\partial F_{j}} = \bigl\langle r \big| \hat{\mu}_j \big| a \bigr\rangle\,.
\end{equation}
Working within the framework of the \gls{cbohf} ansatz, the two terms $A_{ar,bs}$ and $B_{ar,bs}$ contain the standard electron repulsion integrals and the corresponding two-electron \gls{dse} contributions:
\begin{align}
        A_{ar,bs} = &\left(\epsilon_r -\epsilon_a \right) \delta_{rs} \delta_{ab} + \bigl \langle rb  \big| as \bigr\rangle - \bigl \langle rb \big|  sa \bigr\rangle  \label{eq:A1} \\ &+ \sum_c^{N_M} \bigl\langle r \big| \hat{d}_c  \big| a \bigr\rangle \bigl\langle b \big| \hat{d}_c  \big| s \bigr\rangle -  \bigl\langle r \big| \hat{d}_c  \big| s \bigr\rangle \bigl\langle b \big| \hat{d}_c  \big| a \bigr\rangle \label{eq:A2}
\end{align}
and 
\begin{align}
        B_{ar,bs} = & \bigl \langle rs  \big| ab \bigr\rangle - \bigl \langle rs \big|  ba \bigr\rangle \label{eq:B1}  \\ &+ \sum_c^{N_M}  \bigl\langle r \big| \hat{d}_c  \big| s \bigr\rangle \bigl\langle a \big| \hat{d}_c  \big| b \bigr\rangle - \bigl\langle r \big| \hat{d}_c  \big| b \bigr\rangle \bigl\langle s \big| \hat{d}_c  \big| a \bigr\rangle \label{eq:B2}\,.
\end{align}
Here $\epsilon_k$ is the orbital energy of the spin orbital $k$.
Using vector/matrix notations, the \gls{cphf} polarizability $\ev{\bm{\alpha}}_{\mathrm{CBO}}^{\mathrm{cphf}}$ can be written as:
\begin{equation}
\label{eq:alpha_analytic}
    \ev{\alpha_{ij}}_{\mathrm{CBO}}^{\mathrm{cphf}} = 2 \bm{\mu}^T_i \left( \bm{A}+\bm{B} \right)^{-1} \bm{\mu}_j\,,
\end{equation}
where $\bm{\mu}_i$ is the vector of components $ \mu_{i,ar} =  \bigl\langle r \big| \hat{\mu}_i \big| a \bigr\rangle $ and $\left( \bm{A}+\bm{B} \right)$ is the matrix of elements $\left(A_{ar,bs} + B_{ar,bs}\right)$.
In this work we consider two variations of $\ev{\bm{\alpha}}_{\mathrm{CBO}}^{\mathrm{cphf}}$, in one case the \gls{dse} two-electron integrals (Eqs.~\ref{eq:A2} and \ref{eq:B2}) are included and in the second case only the standard electron repulsion integrals are considered (Eqs.~\ref{eq:A1} and \ref{eq:B1}). Note that in both cases, all necessary quantities are obtained from a single \gls{cbohf} calculation. 

\subsection{\gls{cbohf} Raman Activity in the Harmonic Approximation}

To determine Raman activities, i.e. scattering factors, for a molecular system under \gls{vsc}, we extend our recently presented approach for vibro-polaritonic spectra in the harmonic approximation~\cite{Schnappinger2023-wp}. 
The derivative of the polarizability $ \nabla_{k} \bm{\alpha}$ with respect to the $k$th vibro-polaritonic normal mode vector $\bm{Q}^k$ reads
\begin{equation}
   \nabla_{k}  \bm{\alpha}_{ij} =  \sum_{n=1}^{N_A+N_M} \left(\bm{\nabla}\ \  ^{ij}\ev{\bm{\alpha}}_{\mathrm{CBO}} \right)_n \cdot \frac{\bm{Q}^{k}_n}{\sqrt{M_n}}
\end{equation}
where the summation runs over the Cartesian coordinates $N_A$ and photo-displacement coordinates $N_M$. 
The normal mode vector is rescaled by the square root of the corresponding atomic masses, and a mass of one is used for the photonic components.
The necessary polarizability gradient $\bm{\nabla} \ev{\bm{\alpha}}_{\mathrm{CBO}}$ is calculated using finite differences based on the polarizabilities determined numerically or by the liner response formalism, both described in section~\ref{subsec:alpha}.
The elements $ \nabla_{k} \bm{\alpha}_{ij}$ are used to determine the mean polarizability~\cite{Neugebauer2002-nh,Bernath2005-bt} $\bar\alpha_k$ for the vibro-polaritonic normal mode $k$
\begin{equation}
\bar\alpha_k = \left( \nabla_{k}  \bm{\alpha}_{xx} + \nabla_{k}  \bm{\alpha}_{yy} + \nabla_{k}  \bm{\alpha}_{zz}  \right) /3
\end{equation}
as well as the anisotropy $\gamma_k$~\cite{Neugebauer2002-nh,Bernath2005-bt}
\begin{align}
    \gamma_k^2  =  &  \left( \nabla_{k}  \bm{\alpha}_{xx} - \nabla_{k}  \bm{\alpha}_{yy} \right)^2/2 + \left( \nabla_{k}  \bm{\alpha}_{yy} - \nabla_{k}  \bm{\alpha}_{zz} \right)^2/2 +  \left( \nabla_{k}  \bm{\alpha}_{zz} - \nabla_{k}  \bm{\alpha}_{xx} \right)^2/2 
    \nonumber  \\ & + 3 \left( \left(\nabla_{k}  \bm{\alpha}_{xy}\right)^2 +  \left(\nabla_{k}  \bm{\alpha}_{yz}\right)^2  +  \left(\nabla_{k}  \bm{\alpha}_{xz}\right)^2  \right)
\end{align}
Combining both mean polarizability $\bar\alpha_k$ and anisotropy $\gamma_k$ yields the corresponding Raman activity.
The scattering factor $S_k$ for the vibro-polaritonic normal mode $k$ is calculated as follows
\begin{equation}
S_k = 45\bar\alpha_k^2 + 7\gamma_k^2 \,.
\end{equation}

\section{Computational Details}

The calculation of the numerical and \gls{cphf} static polarizability with the \gls{cbohf} ansatz, as well as the calculation of Raman activities, have been implemented in the Psi4NumPy environment~\cite{Smith2018-tu}, which is an extension of the PSI4~\cite{Smith2020-kq} electronic structure package. 
All molecular structures used are based on geometries of a single molecule optimized outside the cavity at the Hartree-Fock level of theory using the aug-cc-pVDZ basis set~\cite{Kendall1992-wu}.
To study the collective effects of \gls{vsc} on the static polarizability and the permanent dipole moment, we use the optimized geometries to create small ensembles by placing $N_{mol}$ replicas of the single molecule separated by \SI{100}{\angstrom} inside the cavity. 
The individual molecular dipole moments and in the case of \ce{CO2} the molecular axes are aligned in parallel to each other. 
These ensembles of $N_{mol}$ molecules are placed at the maximum of the cavity field and are oriented parallel to the polarization axis of a single cavity mode. 
To demonstrate the influence of \gls{vsc} on Raman activities, a single formaldehyde molecule interacting with two orthogonal cavity modes is studied. 
One cavity mode is aligned with the carbonyl group, and the second orthogonal mode is in the molecular plane. 
As shown in the literature~\cite{Liebenthal2024-mx,Schnappinger2024-vt, Huang2025-lu}, the effects on the internal coordinates are small for the coupling strengths studied here; consequently, we do not re-optimize the geometries of the molecular systems in the cavity. 
Molecular reorientation plays an important role, especially in the case of a single cavity mode. 
For linear molecules, the most energetically favorable orientations are those in which the molecules are not coupled to the cavity field at all, that is, the cavity polarization axis is orthogonal to the molecular dipole moment~\cite{Schnappinger2023-hh,Schnappinger2024-vt}. 
However, depending on the coupling strength, the corresponding rotational barriers can be quite small, and to define an upper bound for strong coupling effects, we orient the molecules for maximum coupling to the cavity.
In all \gls{cbohf} calculations performed in this work, we fulfilled the zero transverse electric field condition by minimizing the \gls{cbohf} energy with respect to the photon displacement coordinates~\cite{Schafer2020-cb,Sidler2022-cg,Schnappinger2023-hh}. 
We use an artificially increased coupling strength $\lambda_0$ in the range of \SI{0.001}{\au} to \SI{0.05}{\au}, which corresponds to effective mode volumes, see Eq.~\eqref{eq:coupling}, in the single-molecule case as large as \SI{125.27}{\nano\metre\cubed}~(for $\lambda_0=\SI{0.001}{\au}$) or as small as \SI{1.00}{\nano\metre\cubed}~(for $\lambda_0= \SI{0.05}{\au}$). 
For all spectra shown, the underling signals are broadened by a Lorentzian function with a width of \SI{10}{\per\centi\meter}. All calculations were performed in a reproducible environment using the Nix package manager together with NixOS-QChem \cite{nix} (commit f5dad404) and Nixpkgs (nixpkgs, 22.11, commit 594ef126).

\section{Dipole moment and polarizability under vibrational strong coupling}

First, we investigate how molecular dipole moments and polarizabilities change under \gls{vsc}. 
We limit our discussion in the manuscript to the results for a single \ce{CO} molecule and small ensembles of \ce{CO} molecules. However, to complete the picture, additional results for \ce{LiH}, \ce{CO2}, and \ce{H2O} can be found in section~S1 of the Supporting Information.
Since in the case of \ce{CO} the sign of the molecular dipole moment is wrong at the Hartree-Fock level of theory beyond a minimal basis set~\cite{aszabo82-qc}, we only use the magnitude of the dipole moment vector and not its direction.
To quantify the effect on the static polarizability, we discuss the mean polarizability $\bar{\alpha}$ determined using the eigenvalues of the polarizability tensor $\ev{\bm{\alpha}}_{\mathrm{CBO}}$.

\begin{figure}
     \centering         \includegraphics[width=0.85\textwidth]{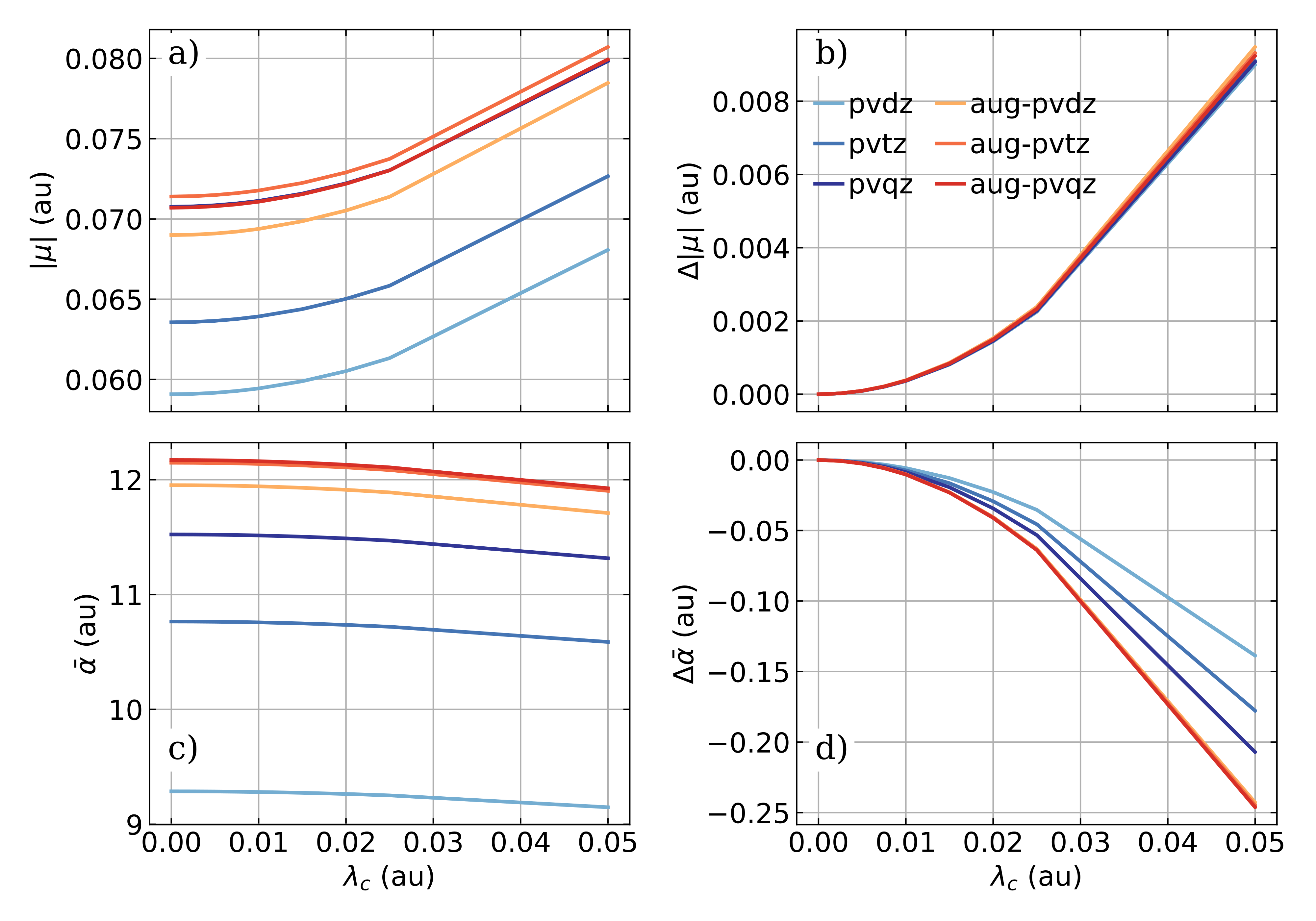}
    \caption{a) Magnitude of the permanent dipole moment $|\mu|$ and b) change in magnitude $\Delta |\mu|$ for a single \ce{CO} molecule as a function of cavity coupling strength $\lambda_c$ for different basis sets (color coded). c) Average polarizability $\bar\alpha$ and d) its change $\Delta \bar\alpha$ for a single \ce{CO} molecule as a function of $\lambda_c$ for different basis sets. The frequency $\omega_c$ of the single cavity mode is resonant with the fundamental transition of the \ce{CO} stretching mode.} 
\label{fig:co_basis}
\end{figure}

Let us first discuss how the size and structure of the basis set affect
the dipole moments and polarizabilities under \gls{vsc}.
For this purpose a single \ce{CO} molecule and the correlation-consistent basis sets, developed by Dunning and coworkers~\cite{Dunning1989-xc,Kendall1992-wu}, are used in the standard and augmented versions up to quadruple-zeta size.
The magnitude of the dipole moment $|\mu|$ and the mean polarizability $\bar{\alpha}$ (determined with $\ev{\bm{\alpha}}_{\mathrm{CBO}}^{\mathrm{num}}$) as well as their difference $\Delta|\mu|$ and $\Delta \bar{\alpha}$ with respect to the field-free case are shown in Fig.~\ref{fig:co_basis} for different basis sets as a function of $\lambda_c$.
The values $\ev{\bm{\alpha}}_{\mathrm{CBO}}^{\mathrm{num}}$ are determined using a field strength of \SI{0.00001}{\au}.

Qualitatively, the same trends of $|\mu|$ (Fig.~\ref{fig:co_basis}~a) and $\bar\alpha$ (Fig.~\ref{fig:co_basis}~c) are observed for all basis sets with increasing $\lambda_c$.
The dipole moment increases slightly and the polarizability decreases in agreement with our previous work~\cite{Schnappinger2024-vt}. 
The same general trends are observed for both molecular properties in the case of \ce{LiH}, \ce{CO2}, and \ce{H2O}, which are shown in Figs.~S1,~S2,~and~S3 in the Supporting Information. 
The absolute values of $|\mu|$ and $\bar\alpha$ converge with increasing basis set size.
In general, the values for the augmented basis sets are more similar and converge significantly faster for $\bar\alpha$. 
Consequently, it seems reasonable that at least aug-cc-pVDZ should be used to determine both $|\mu|$ and $\bar\alpha$, since it partially outperforms the standard correlation-consistent basis sets.
The effect of \gls{vsc} on the dipole moment is visualized as $\Delta |\mu|$ in Fig.~\ref{fig:co_basis}~b) and shows very little dependence on the chosen basis set.
Only for a coupling strength $\lambda_c$ greater than 0.03 some differences between the basis sets can be noticed.
In the case of $\Delta \bar\alpha$, the dependence on the base set is more pronounced and the main difference is observed between the standard and the augmented versions.
The augmented basis sets describe the effect of \gls{vsc} on the polarizability almost identically, while the non-augmented basis sets versions underestimate the effect but tend to converge to the same result with increasing size. 
Note that for \ce{LiH} and \ce{H2O} the dependence of $|\mu|$ on the basis set is more pronounced, since the corresponding dipole moments are per se larger than in the case of \ce{CO}.

\begin{figure}
     \centering
    \includegraphics[width=0.9\textwidth]{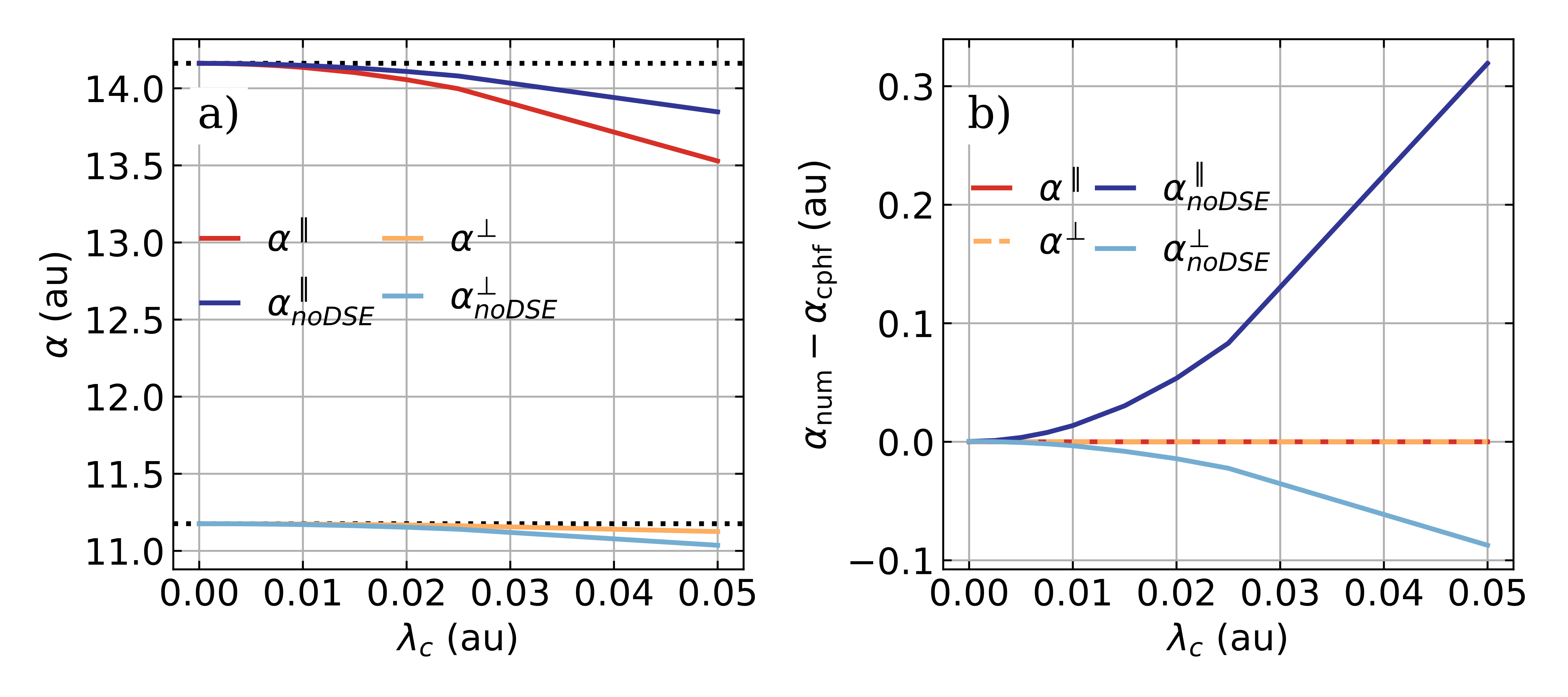}
    \caption{a) The two non-degenerate eigenvalues of the polarizability tensor $\ev{\bm{\alpha}}_{\mathrm{CBO}}^{\mathrm{cphf}}$ of \ce{CO} as a function of the cavity coupling strength $\lambda_c$ once calculated including the \gls{dse} two-electron contributions (red and yellow) and once without it (dark and light blue). The black dashed lines represent the corresponding field-free eigenvalues. b) The difference between the \gls{cphf} polarizability eigenvalues (with and without \gls{dse} two-electron contributions) and the $\ev{\bm{\alpha}}_{\mathrm{CBO}}^{\mathrm{num}}$  eigenvalues. The  frequency $\omega_c$ of the single cavity mode is resonant with the fundamental transition of the \ce{CO} stretching mode. All values calculated on the CBO-HF/aug-cc-pVQZ level of theory.} 
\label{fig:co_method}
\end{figure}

To validate the \gls{cphf} implementation of the polarizability $\ev{\bm{\alpha}}_{\mathrm{CBO}}^{\mathrm{cphf}}$, we compare the two non-degenerate eigenvalues of the polarizability tensor with and without the \gls{dse} two-electron integrals (Eqs.~\ref{eq:A2} and \ref{eq:B2}) with respect to the formally exact $\ev{\bm{\alpha}}_{\mathrm{CBO}}^{\mathrm{num}}$ values. 
This comparison is shown in Fig.~\ref{fig:co_method} for the case of a single \ce{CO} molecule resonantly coupled to a single cavity mode. 
The values $\ev{\bm{\alpha}}_{\mathrm{CBO}}^{\mathrm{num}}$ are determined using a field strength of \SI{0.00001}{\au}.

Regardless of whether the \gls{dse} two-electron contributions are included or not, the same trends are observed for the larger eigenvalue $\alpha^{\parallel}$ and the smaller degenerate eigenvalues $\alpha^{\bot}$ of the polarizability tensor with increasing coupling strength, see Fig.~\ref{fig:co_method}~a).
Both eigenvalues of $\ev{\bm{\alpha}}_{\mathrm{CBO}}$ decrease with increasing $\lambda_c$.
In our simulations, the principal axis of $\alpha^{\parallel}$ is aligned with the molecular axis and the cavity polarization axis. 
Consequently, $\alpha^{\parallel}$ is more affected by the coupling to the cavity mode than the two degenerate eigenvalues $\alpha^{\bot}$. 
Interestingly, the effect of the \gls{dse} contribution is different for $\alpha^{\parallel}$ and $\alpha^{\bot}$.
For $\alpha^{\parallel}$ the decrease is stronger with \gls{dse} contributions included, while for $\alpha^{\bot}$ the opposite effect is observed. 
We can report similar effects for \ce{LiH}, but not for \ce{CO2}  as shown in Figs.~S4 and~S5 in the Supporting Information.
The comparison of the two types of $\ev{\bm{\alpha}}_{\mathrm{CBO}}^{\mathrm{cphf}}$ eigenvalues with the corresponding numerically determined eigenvalues is shown in Fig.~\ref{fig:co_method}~b).
The $\ev{\bm{\alpha}}_{\mathrm{CBO}}^{\mathrm{cphf}}$ values including the \gls{dse} contributions (red and yellow lines) are identical to the $\ev{\bm{\alpha}}_{\mathrm{CBO}}^{\mathrm{num}}$ values. 
Neglecting the \gls{dse} contributions (blue lines) leads to a deviation from the numerical results with increasing coupling strength.

\begin{figure}
     \centering
    \includegraphics[width=0.9\textwidth]{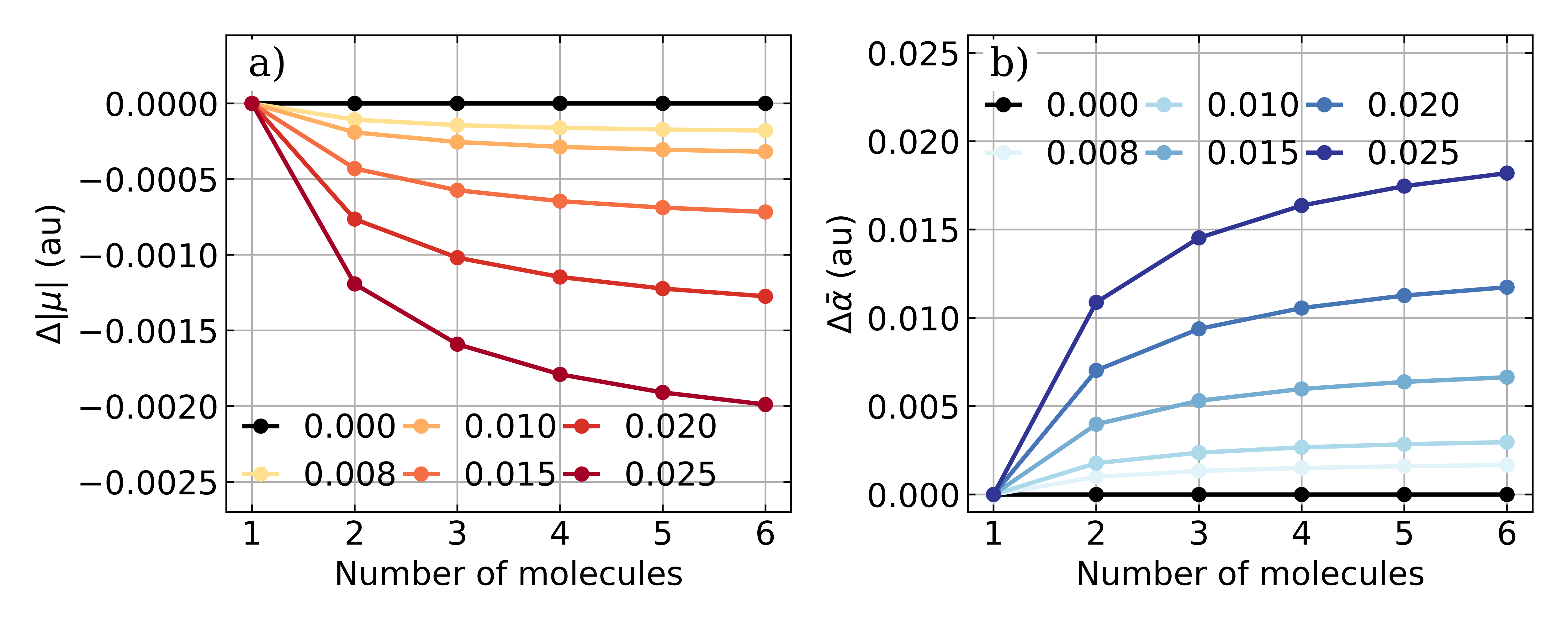}
    \caption{a) Change in the dipole moment per molecule $\Delta \left|\mu\left(N_{mol}, \lambda_0 \right)\right|$ and b) change in the mean polarizability per molecule $\Delta \bar{\alpha}\left(N_{mol}, \lambda_0 \right)$ as function of the number of \ce{CO} molecules for different values of $\lambda_0$ (color coded). The frequency $\omega_c$ of the single cavity mode is resonant with the fundamental transition of the \ce{CO} stretching mode. All values calculated on the CBO-HF/aug-cc-pVDZ level of theory.} 
\label{fig:co_scaling}
\end{figure}

We now turn to possible collective effects on these properties within small molecular ensembles. 
In a recent publication by Horak et al.~\cite{Horak2024-fa} a local change in molecular polarizability induced by collective strong coupling was observed in a simple harmonic model. 
In order to determine such a local effect on both the dipole moment and the polarizability, we calculate the changes per molecule as a function of the number of molecules and the coupling strength $\lambda_0$:
\begin{align}
        \Delta \left|\mu\left(N_{mol}, \lambda_0 \right)\right| &  = \frac{\left|\mu_{N_{mol}} \left( \lambda_c \right)\right|}{N_{mol}} - \left|\mu_{1} \left( \lambda_0 \right)\right| \\
        \Delta \bar{\alpha}\left(N_{mol}, \lambda_0 \right) & = \frac{\bar{\alpha}_{N_{mol}} \left( \lambda_c \right)}{N_{mol}} - \bar{\alpha}_{1} \left( \lambda_0 \right) 
\end{align}
Note that the collective coupling strength $\lambda_c$ for the molecular ensembles is kept constant by the rescaling of $\lambda_0$ (see Eq.~\ref{eq:coupling}), and the local coupling per molecule decreases as the number of molecules increases. 
In a situation without cavity-induced collective effects, the dipole moment and the mean polarizability per molecule would be the same as in the single molecule case, $\left|\mu_{1} \left( \lambda_0 \right)\right|$ and $\bar{\alpha}_{1} \left( \lambda_0 \right)$, due to the rescaling of the collective coupling strength.
Consequently $\Delta \left|\mu\left(N_{mol}, \lambda_0 \right)\right|$ and $\Delta \bar{\alpha}\left(N_{mol}, \lambda_0 \right)$ would be exactly zero. 
Fig.~\ref{fig:co_scaling} shows the change in the dipole moment per molecule $\Delta \left|\mu\left(N_{mol}, \lambda_0 \right)\right|$ and the change in the mean polarizability per molecule $\Delta \bar{\alpha}\left(N_{mol}, \lambda_0 \right)$ for different values of $\lambda_0$ as a function of the number of \ce{CO} molecules.
Additional results for \ce{LiH}, \ce{CO2}, and \ce{H2O} are shown in the supporting informations Figs.~S6,~S7, and~S8.

As shown in Fig.~\ref{fig:co_scaling} both $\Delta \left|\mu\left(N_{mol}, \lambda_0 \right)\right|$ and $\Delta \bar{\alpha}\left(N_{mol}, \lambda_0 \right)$ are nonzero in our simulations, regardless of the chosen coupling strength $\lambda_0$.
The change in the dipole moment per molecule $\Delta \left|\mu\right|$ (Fig.~\ref{fig:co_scaling}~a)) decreases with increasing number of molecules in the ensemble and approaches a constant nonzero value that decreases with increasing $\lambda_0$.
The collective interaction via the cavity mode reduces the change in dipole moment expected from the single molecule case.
Interestingly, the exact same behavior can be found for all molecules studied in this work, see Figs.~S6,~S7,~and~S8 in the Supporting Information, indicating that this collective effect is rather general.
Note that the only exception is \ce{CO2}, which has no permanent dipole moment and the coupling strength used is not large enough to induce a relevant dipole moment in the system. 
The change in the mean polarizability per molecule $\Delta \bar{\alpha}$, shown in Fig~\ref{fig:co_scaling}~b), increases with the number of molecules in the ensemble and approaches a constant non-zero value that increases with increasing $\lambda_0$. 
This result indicates that cavity-mediated collective interactions increase the polarizability per molecule in comparison to the single-molecule case. 
Again, this trend seems quite general, as it is also observed for \ce{LiH}, \ce{CO2}, and \ce{H2O}, as shown in the Supporting Information Figs.~S6,~S7,~and~S8. 
For all molecules studied, the change per molecule for both properties 
the scaling with respect to $N_{mol}$ is proportional to $1-\sfrac{1}{ N_{mol}}$ and approaches a constant, non-zero value that depends on the coupling strength. 
We had already observed the same scaling behavior for the \gls{dse} induced intermolecular dipole-dipole energy contribution for small ensembles of hydrogen fluoride molecules~\cite{Schnappinger2023-hh}. 
We consider the fact that we observe a collective effect on both dipole moment and polarizability under \gls{vsc} to be one of the central findings of this work, in line with the results obtained previously for an analytic harmonic model~\cite{Horak2024-fa}. 
These collective effects clearly indicate that a single molecule in the cavity-coupled ensemble is not independent from the rest, even if the corresponding modifications are small. 

\section{Vibro-Polaritonic Spectra for Formaldehyde}

\begin{figure}
     \centering
         \includegraphics[width=0.75\textwidth]{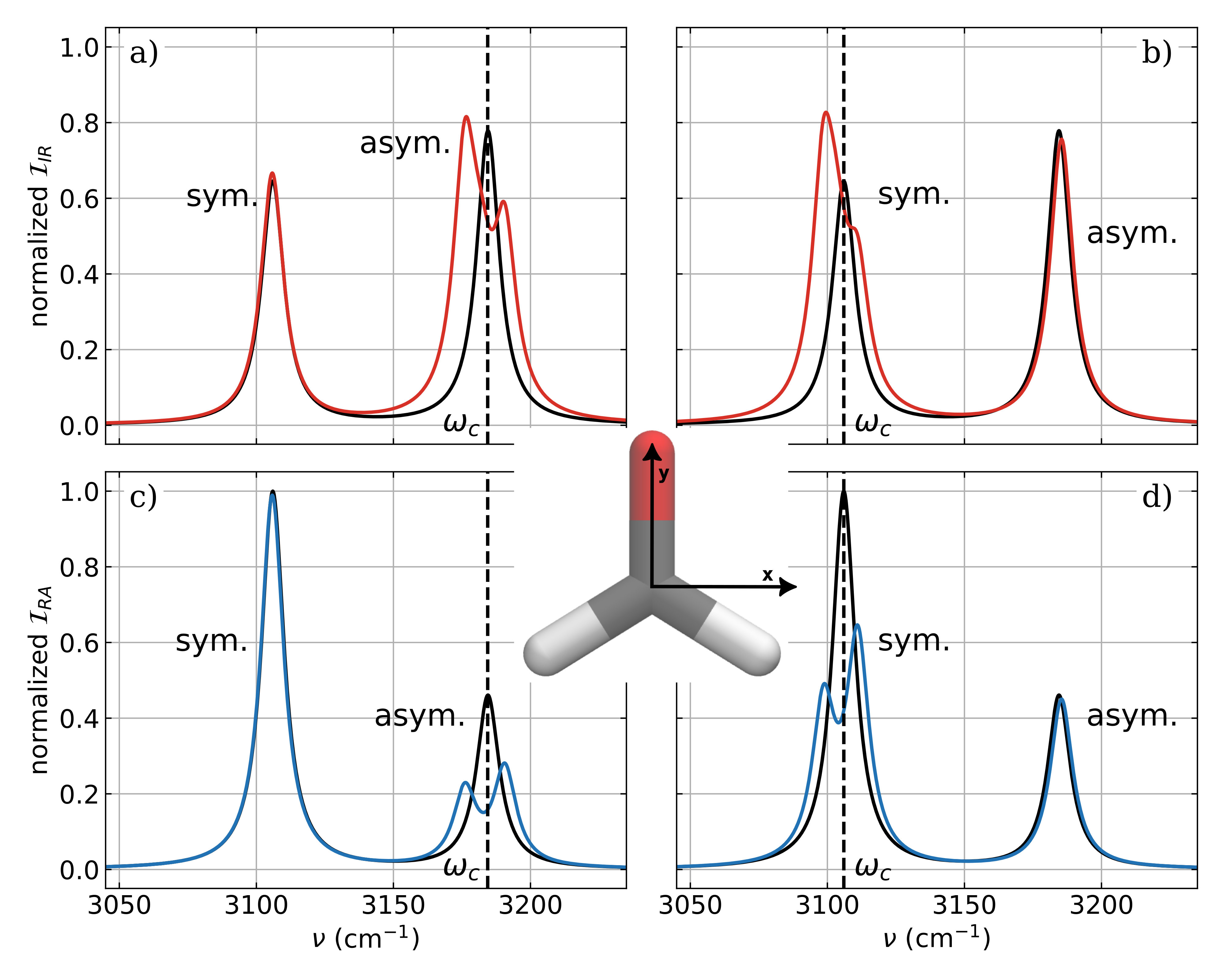}
    \caption{a) and b) vibro-polaritonic IR spectra shown in red and c) and d) vibro-polaritonic Raman spectra shown in blue of a single formaldehyde molecule coupled to two cavity modes.
    Only the frequency range above \SI{3000}{\per\centi\meter} for the symmetric and asymmetric \ce{CH2} stretching modes is shown. 
    For a) and c) the cavity modes are resonant with the asymmetric stretching and for b) and d) with the symmetric stretching, the corresponding frequencies $\omega_c$ are marked with black dashed lines.
   The spectra are normalized with respect to the maximum intensity of the cavity-free spectrum, shown in black. The molecular orientation with respect to the two cavity polarization axes $x$ and $y$ is shown in the middle inset. All spectra are calculated at the CBO-HF/aug-cc-pVDZ level of theory using a coupling strength $\lambda_{c}$ of \SI{0.010}{\au}.} 
\label{fig:form_resonant}
\end{figure}

Apart from being an important molecular property for understanding molecular interactions such as the London dispersion forces~\cite{London1937-um,Maitland1983IntermolecularFT}, the polarizability also determines how molecules interact with light.
In this work, we combine the static polarizability with our work on vibro-polaritonic spectra in the harmonic approximation~\cite{Schnappinger2023-wp} to calculate Raman activity and scattering factors under \gls{vsc}. 
As a first demonstration and also to investigate the different effects of \gls{vsc} on IR and Raman spectra, we simulate both vibro-polaritonic spectra of a single formaldehyde molecule coupled to two cavity modes of orthogonal polarization with the same frequency, effectively modeling a simplified Fabry-P\'erot-like setup. 
The single formaldehyde molecule is oriented with respect to the center of the nuclear charges, the molecular plane is in the $x$-$y$ plane, and the carbonyl group is aligned with the $y$ axis of the laboratory frame, as shown in the inset of Fig.~\ref{fig:form_resonant}.
The polarization axes of the two cavity modes are aligned with the $x$ axis and the $y$ axis, respectively. 
Formaldehyde has six vibrational degrees of freedom: three bending modes, one out of plane and two in plane, the \ce{CO} stretching mode and one symmetric and one asymmetric \ce{CH2} stretching mode.
The symmetric and asymmetric \ce{CH2} stretching modes are both IR and Raman active, and are reasonably close in energy with $\Delta \nu \approx \SI{80}{\per\centi\meter}$.
The transition dipole moment for the symmetric mode points along the $y$ axis and for the asymmetric mode along the $x$ axis.
Therefore, we choose the cavity frequencies $\omega_c$ to be larger than \SI{3000}{\per\centi\meter} to focus the discussion on the spectral range of these two stretching modes.
The bare molecular vibronic IR and Ramana spectra and the vibro-polaritonic IR and Ramana spectra of a single formaldehyde molecule coupled to two cavity modes are shown in Fig.~\ref{fig:form_resonant} for two different cavity frequencies. 

Let us start by analyzing the vibro-polaritonic IR spectra shown in Fig.~\ref{fig:form_resonant}~a) and~b). 
The asymmetric stretching mode ($\nu = \SI{3184}{\per\centi\meter}$ HF/aug-cc-pVDZ level of theory) is more intense than the symmetric stretching mode ($\nu = \SI{3106}{\per\centi\meter}$ HF/aug-cc-pVDZ level of theory) for the spectrum of the uncoupled formaldehyde molecule. 
As expected, the splitting into a \gls{lp} and a \gls{up} transition is observed for the symmetric/asymmetric stretching mode when coupled resonantly to the optical cavity.
In both cases, the \gls{lp} transition is broadened by the presence of the second cavity mode; the reason for this will be discussed later. 
In the case of resonant coupling to the symmetric stretching mode (Fig.~\ref{fig:form_resonant}~b)) this leads to the situation that the \gls{up} transition is only visible as a shoulder for the chosen coupling strength and broadening of the spectrum.
The combined intensity of the \gls{lp} and \gls{up} transition is larger than the intensity of the uncoupled molecular transition, which is reasonable given our observation that the dipole moment itself increases under \gls{vsc}.
For the chosen coupling strength of \SI{0.010}{\au}, the non-resonant transition is not affected in terms of both intensity and frequency. 
In agreement with the literature~\cite{Sidler2024-tw,Schnappinger2023-wp,Fischer2024-mk,Fiechter2024-gl,Huang2025-lu} the formed pair of vibro-polaritonic transition is asymmetric in two ways. 
First, the Rabi splitting $\Omega_R$ between \gls{lp} and \gls{up} is asymmetric, where \gls{lp} is more red shifted than \gls{up} is blue shifted with respect to $\omega_c$.
This asymmetry can be understood as a detuning and a change in optical length of the cavity when interacting with the molecule~\cite{Sidler2024-tw,Schnappinger2023-wp,Fiechter2024-gl}.
Second, the signal intensity is asymmetric, with the \gls{lp} transition more intense than the \gls{up} transition, which has also been observed in the literature~\cite{Fischer2024-mk,Huang2025-lu}. 
Recently, Huang and Liang~\cite{Huang2025-lu} explained this asymmetry based on a simplified Hessian model. 
They argue that the transition dipole moment of the \gls{lp} state is formed by the positive linear combination of the dipole outside the cavity and a cavity-induced dipole moment, while the corresponding \gls{up} dipole moment is formed by the negative linear combination, leading to a higher intensity for the \gls{lp} transition and a lower intensity for the \gls{up} transition.

Since both vibro-polaritonic IR spectra and Raman spectra are based on the same Hessian matrix, we only discuss the Raman intensities of the vibro-polaritonic transitions shown in Fig.~\ref{fig:form_resonant}~c) and~d).
In the case of uncoupled formaldehyde, the symmetric stretching mode ($\nu = \SI{3106}{\per\centi\meter}$) is more intense than the asymmetric stretching mode ($\nu = \SI{3184}{\per\centi\meter}$). 
When resonantly coupled to the cavity modes, we can also observe the splitting into a \gls{lp} and a \gls{up} transition in the vibro-polaritonic Raman spectra. 
Note that this is only possible because both \ce{CH2} stretching modes are IR and Raman active; otherwise, there would be no formation of polaritonic states, i.e. the transition is not IR active, or they would not be observable in the Raman spectrum.
In contrast to the corresponding IR spectra, the Raman intensities of the \gls{lp} and \gls{up} transitions are significantly lower than for the bare molecular transition.
This can be related to our observation that the polarizability itself decreases under \gls{vsc}.
In agreement with Huang and Liang~\cite{Huang2025-lu} we observe an asymmetry in the intensities of the \gls{lp} and \gls{up} transitions. In contrast to the vibro-polaritonic IR spectra, the \gls{up} transition is more intense compared to the \gls{lp} transition in the Raman spectra. 

\begin{figure}
     \centering
         \includegraphics[width=0.9\textwidth]{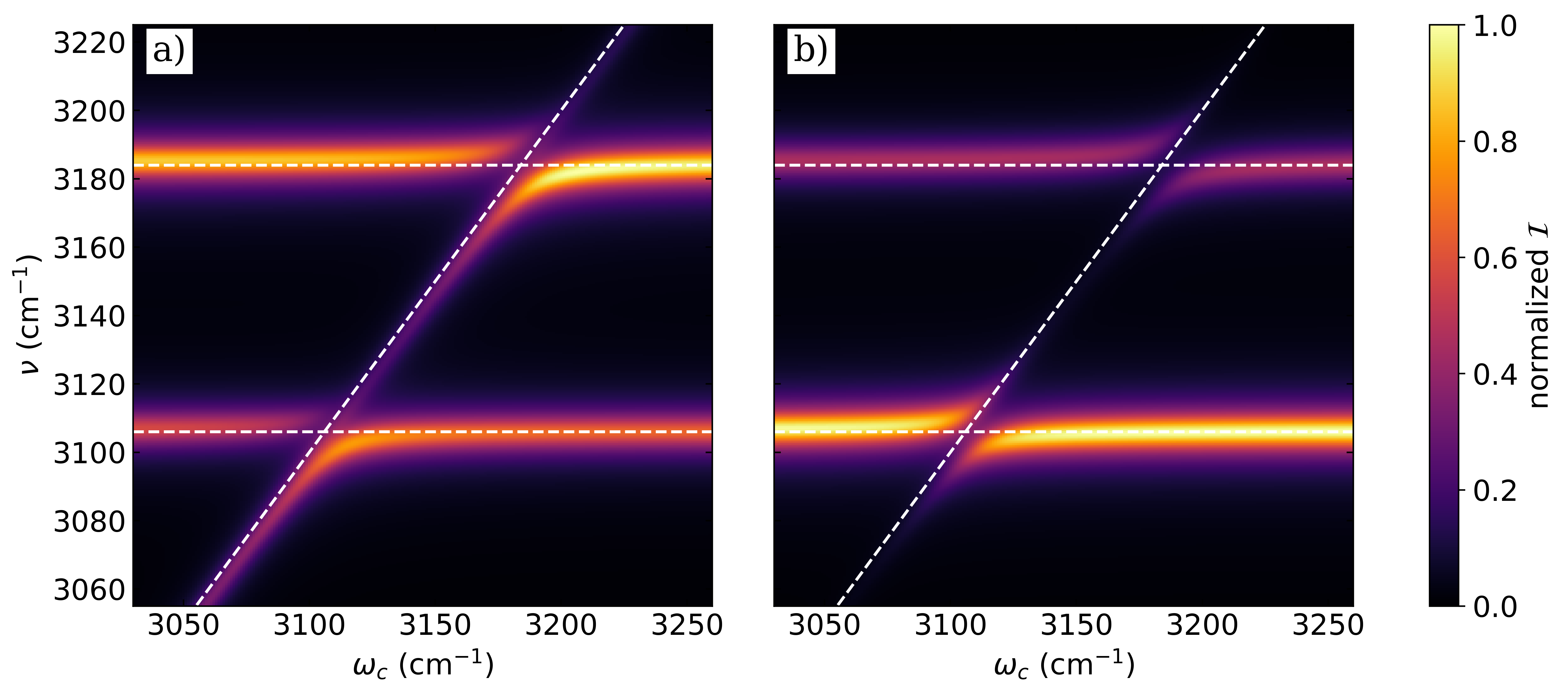}
    \caption{Normalized dispersion with respect to cavity frequency $\omega_c$ of a) the vibro-polaritonic IR spectrum and b) the vibro-polaritonic Raman spectrum of a single formaldehyde molecule for different cavity frequencies.
    The horizontal white dashed lines indicate the bare molecule transitions and the diagonal white dashed line indicates the bare cavity frequency. All underlying spectra are calculated at the CBO-HF/aug-cc-pVDZ level of theory using a coupling strength $\lambda_{c}$ of \SI{0.010}{\au}.} 
\label{fig:form_dispersion}
\end{figure}

In Fig.~\ref{fig:form_dispersion} the dispersion with respect to cavity frequency $\omega_c$ for both the vibro-polaritonic IR spectrum and the vibro-polaritonic Raman spectrum is shown using the same parameters as previously used in Fig.~\ref{fig:form_resonant}. The dispersion curves are obtained by scanning $\omega_c$ for both cavity modes in the range from \SI{3030}{\per\centi\meter} to \SI{3260}{\per\centi\meter} with a step size of \SI{0.5}{\per\centi\meter}.

In the vicinity of the two resonances (around \SI{3106}{\per\centi\meter} and \SI{3184}{\per\centi\meter}, i.e. close to zero detuning of the cavity), the formation of the \gls{lp} and the \gls{up} leads to the appearance of the well-known avoided crossing pattern in the dispersion curves of both spectra, shown in Fig.~\ref{fig:form_resonant}~a) and b).
As detuning increases to smaller or larger values of $\omega_c$, the observed vibro-polaritonic transitions become more molecular or photonic. 
The more photonic transition loses intensity, while the more molecular transition remains visible and constant in frequency. 
When the dispersion curves of the IR and Raman spectra are compared, the more photonic transition loses its Raman intensity much faster than its IR intensity. 
Another difference between the IR and Raman signals is observed at larger detunings. 
A direct comparison of the IR intensities at $\omega_c =\SI{3030}{\per\centi\meter}$ (far left)  and $\omega_c =\SI{3260}{\per\centi\meter}$ (far right) for the signal corresponding to the symmetric stretching mode ($\nu = \SI{3180}{\per\centi\meter}$) clearly shows that the intensity increases for larger $\omega_c$, even though the transition is no longer resonantly coupled. 
In contrast, the Raman intensities of the out-of-resonance molecular transitions remain approximately the same before and after hybridization. 
In addition, the discussed decoupling of molecular and photonic transitions is also clearly visible in the $|a_c|^{2}$ values shown in Fig.~S9 of the Supporting Information. 
In the \gls{cboa} the normal mode vectors have terms $a_c$ describing the change in the classical photon displacement field coordinates $q_c^{(x)}$ and $q_c^{(y)}$. 
The value of $|a_c|^{2}$ for a given normal mode is a measure of how strongly the corresponding vibrational transition interacts with the photon field and allow us to estimate the photonic character of the corresponding transitions~\cite{Schnappinger2023-wp}.

\begin{figure}[]
     \centering
         \includegraphics[width=0.7\textwidth]{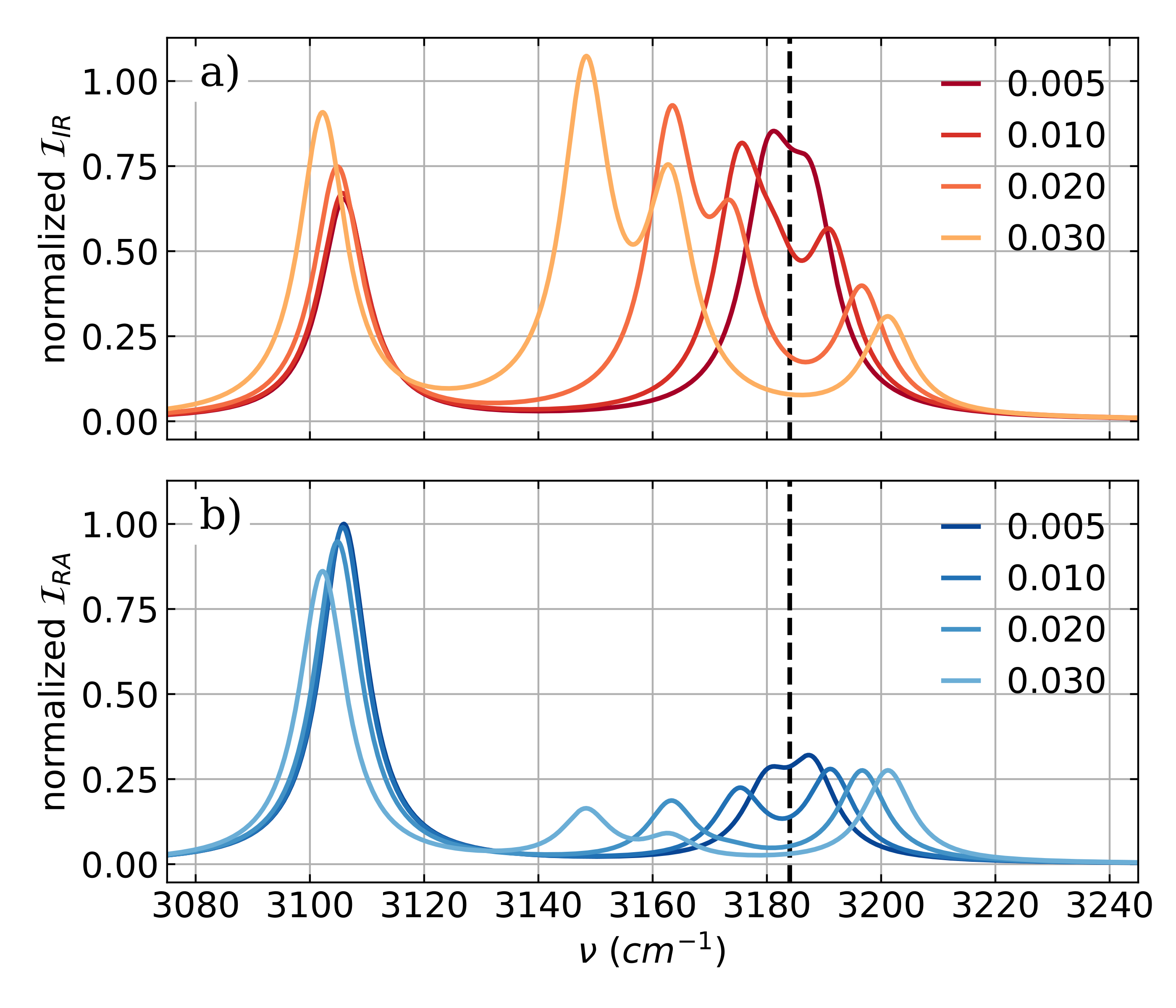}
    \caption{a) vibro-polaritonic IR spectra and b) vibro-polaritonic Raman spectra of a single formaldehyde molecule for increasing coupling strength $\lambda_{c}$. The cavity frequency is chosen to be resonant with  resonant with the asymmetric \ce{CH2} stretching mode of of \SI{3184}{\per\centi\meter} and  indicated as black dashed line.} 
\label{fig:scan_lam3}
\end{figure}

\begin{figure}
     \centering
         \includegraphics[width=0.75\textwidth]{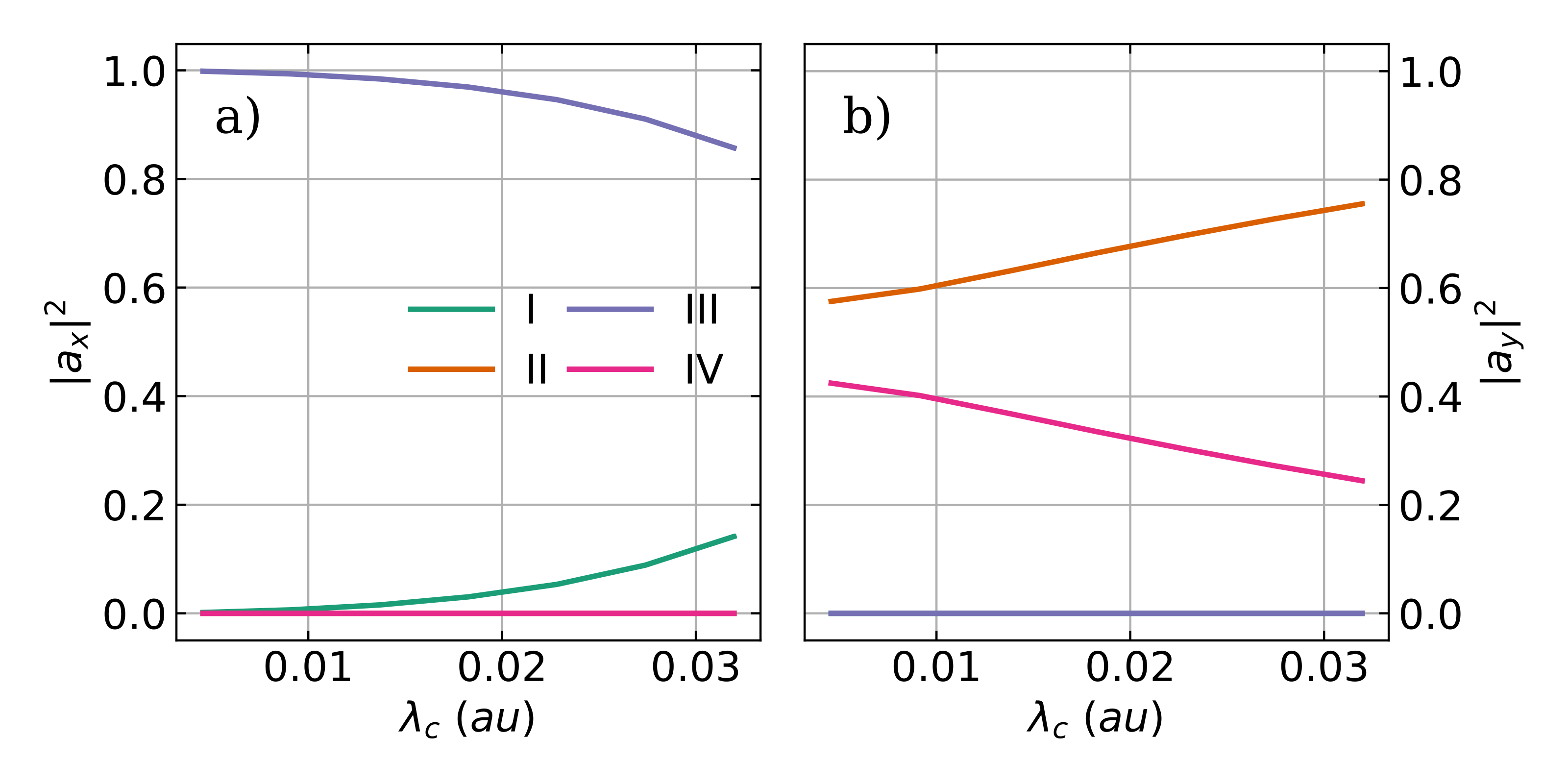}
    \caption{$|a_c|^{2}$ values of the four relevant vibro-polaritonic transitions (sorted by ascending frequency and labeled with Latin numbers) corresponding to a) the cavity mode with $x$ polarization and b) the cavity mode with $y$ polarization as a function of $\lambda_c$. The underlying normal modes are calculated at the CBO-HF/aug-cc-pVDZ level of theory and the coupling strength $\lambda_{c}$ increases from \SI{0.005}{\au} to \SI{0.033}{\au}. The cavity frequency $\omega_c$ of \SI{3184}{\per\centi\meter} resonates with the asymmetric stretching mode.} 
\label{fig:coef_lam3}
\end{figure}

The last aspect we want to discuss is how both vibro-polaritonic IR spectra and Raman spectra change with increasing coupling strength $\lambda_c$ for the case of a single formaldehyde molecule coupled to two cavity modes resonant with the asymmetric \ce{CH2} stretching mode. 
The spectra for different values of $\lambda_c$ are shown in Fig~\ref{fig:scan_lam3} and the $|a_c|^{2}$ values for all four relevant transitions (sorted by ascending frequency and labeled with Latin numbers) as function of $\lambda_c$ are shown in Fig.~\ref{fig:coef_lam3}.
For $\lambda_c = \SI{0.005}{\au}$ the light-matter coupling is strong enough to observe the formation of the \gls{lp} transition and the \gls{up} transition, and two strongly overlapping signals are visible.
As discussed above, the \gls{lp} transition is more intense in the IR spectrum and the \gls{up} transition is more intense in the Raman spectrum. 
The pair \gls{lp} and \gls{up} is formed as expected by the asymmetric stretching mode and the cavity mode with the $y$ polarization axis, see Fig.~\ref{fig:coef_lam3}~b) where both transitions are labeled II and IV. 
The non-resonant symmetric stretching mode is not altered by the cavity interaction.
As the coupling strength increases, the Rabi splitting $\Omega_R$ between the II and IV transitions increases. The \gls{up} transition (IV, highest frequency signal) is blue shifted and loses IR intensity.
In contrast, its Raman intensity remains constant with increasing $\lambda_c$.
The corresponding \gls{lp} transition (II) is more red shifted and gains IR intensity while losing Raman intensity. 
At the same time, an additional transition becomes more pronounced in the IR spectrum between the \gls{lp} and \gls{up} transitions, starting as a shoulder of the \gls{lp} signal for $\lambda_c = \SI{0.010}{\au}$.
In the Raman spectra this transition is only observable as a very weak signal around \SI{3160}{\per\centi\meter} for the highest coupling strength studied, see Fig~\ref{fig:scan_lam3}~b).
Analyzing the corresponding $|a_c|^{2}$ values shown in Fig.~\ref{fig:coef_lam3}, it becomes clear that this transition originates from the cavity photon mode with the $y$ polarization axis, labeled III. This transition is mostly photonic for all coupling strengths. 
However, with increasing coupling strength, this cavity photon mode can weakly hybridize with the off-resonant symmetric stretching mode, labeled I in Fig.~\ref{fig:coef_lam3}, leading to a strong redshift of the cavity mode and an increase in intensities. 
This mixing also affects the symmetric stretching mode for coupling strengths greater than \SI{0.020}{\au}. 
The mostly molecular mode is also slightly red-shifted, while gaining IR intensity and losing Raman intensity. 
In summary, for the higher coupling strengths we observe some kinds of nested polaritons which can be understood as two pairs of \gls{lp} and \gls{up} transitions. 
These nested polariton features due to interactions with off-resonant cavity modes are also observed in \gls{vsc} experiments~\cite{George2016-sy,Wright2023-xx} in the strong and ultra-strong coupling regime.
However, it should be noted that both experiments are not fully comparable, since they are based on collective strong coupling in the presence of multiple (more than two) cavity modes. 

\section{Summary and Conclusion}
We have derived an analytic formulation of the static polarizability for the cavity Born-Oppenheimer Hartree-Fock ansatz~\cite{Schnappinger2023-hh} 
using linear-response theory. 
The obtained polarizabilities have been used to calculate the vibro-polaritonic Raman spectra under \gls{vsc} in the harmonic approximation using a wave-function based methodology.
By studying the effect of \gls{vsc} for different molecules (\ce{CO}, \ce{LiH}, \ce{CO2} and \ce{H2O}) in the single molecule case, we were able to generalize our previous observations~\cite{Schnappinger2024-vt}
that the permanent dipole moment increases while the static polarizability decreases with increasing coupling strength $\lambda_c$. 
These two trends were observed for all four molecules studied, but for the case of \ce{CO2} without a permanent dipole moment, \gls{vsc} does not induce a significant dipole moment for the coupling strength used in this work. 
One of our main objectives in this work was to investigate whether a cavity mediated local modification of $\bm{\alpha}$ under \gls{vsc}, as demonstrated for a full-harmonic model~\cite{Horak2024-fa}, exists when describing real molecular ensembles. 
By calculating the mean polarizability per molecule for ensembles of \ce{CO}, \ce{LiH}, \ce{CO2}, and \ce{H2O} molecules under \gls{vsc}, we observed a cavity-mediated local modification.
The change per molecule is proportional to $1-\sfrac{1}{ N_{mol}}$ and approaches a constant, non-zero value that depends on the coupling strength. 
We had already observed the same scaling behavior for the \gls{dse} induced intermolecular dipole-dipole energy contribution~\cite{Schnappinger2023-hh}. 
A similar trend is observed for the change in the magnitude of the permanent dipole moment per molecule in such ensembles. 
These per-molecule changes in both molecular properties can be interpreted as a situation in which the single molecule in the cavity-coupled ensemble is no longer independent of the rest. 

When simulating Raman spectra using Raman activities based on the harmonic approximation for a single formaldehyde molecule coupled to two cavity modes, we observed similar effects of \gls{vsc} as Huang and Liang~\cite{Huang2024-is,Huang2025-lu}.
The Raman intensities for \gls{lp} and \gls{up} are asymmetric, but otherwise behave differently compared to the corresponding IR intensities. 
The Raman signal of the \gls{up} transition is more intense than the \gls{lp} signal, and overall the signals weaken with increasing coupling strength.
Although these trends are present in our theoretical description, we are unable to provide a comprehensive explanation, and a more in-depth study of vibro-polaritonic Raman spectra is needed. 
However, we would like to emphasize that due to the Fabry-P\'erot-like setup and the chosen molecular system, we are able to observe nested polariton features in the IR and Raman spectra due to interactions with off-resonant cavity modes, which are also observed in \gls{vsc} experiments~\cite{George2016-sy,Wright2023-xx}. 

\begin{acknowledgement}
The authors thank Dominik Sidler, Eric Fischer and Michael Ruggenthaler for inspiring discussions and helpful comments. This project has received funding from the European Research Council (ERC) under the European Union’s Horizon 2020 research and innovation program (grant agreement no. 852286). Support from the Swedish Research Council (Grant No.~VR 2024-04299) is acknowledged.
\end{acknowledgement}

\begin{suppinfo}
See the supplementary material for additional figures showing the dipole moments and polarizabilities of \ce{LiH}, \ce{CO2}, and \ce{H2O} under vibrational strong  coupling and the vibro-polaritonic normal mode analysis for formaldehyde. All data underlying this study are available from the corresponding author upon reasonable request.
\end{suppinfo}

\bibliography{lit.bib}

\begin{mcitethebibliography}{66}
\providecommand*\natexlab[1]{#1}
\providecommand*\mciteSetBstSublistMode[1]{}
\providecommand*\mciteSetBstMaxWidthForm[2]{}
\providecommand*\mciteBstWouldAddEndPuncttrue
  {\def\EndOfBibitem{\unskip.}}
\providecommand*\mciteBstWouldAddEndPunctfalse
  {\let\EndOfBibitem\relax}
\providecommand*\mciteSetBstMidEndSepPunct[3]{}
\providecommand*\mciteSetBstSublistLabelBeginEnd[3]{}
\providecommand*\EndOfBibitem{}
\mciteSetBstSublistMode{f}
\mciteSetBstMaxWidthForm{subitem}{(\alph{mcitesubitemcount})}
\mciteSetBstSublistLabelBeginEnd
  {\mcitemaxwidthsubitemform\space}
  {\relax}
  {\relax}

\bibitem[Fregoni \latin{et~al.}(2022)Fregoni, Garcia-Vidal, and
  Feist]{Fregoni2022-op}
Fregoni,~J.; Garcia-Vidal,~F.~J.; Feist,~J. Theoretical Challenges in
  Polaritonic Chemistry. \emph{ACS Photonics} \textbf{2022}, \emph{9},
  1096--1107\relax
\mciteBstWouldAddEndPuncttrue
\mciteSetBstMidEndSepPunct{\mcitedefaultmidpunct}
{\mcitedefaultendpunct}{\mcitedefaultseppunct}\relax
\EndOfBibitem
\bibitem[Dunkelberger \latin{et~al.}(2022)Dunkelberger, Simpkins, Vurgaftman,
  and Owrutsky]{Dunkelberger2022-oh}
Dunkelberger,~A.~D.; Simpkins,~B.~S.; Vurgaftman,~I.; Owrutsky,~J.~C.
  {Vibration-Cavity} Polariton Chemistry and Dynamics. \emph{Annu. Rev. Phys.
  Chem.} \textbf{2022}, \emph{73}, 429--451\relax
\mciteBstWouldAddEndPuncttrue
\mciteSetBstMidEndSepPunct{\mcitedefaultmidpunct}
{\mcitedefaultendpunct}{\mcitedefaultseppunct}\relax
\EndOfBibitem
\bibitem[Ebbesen \latin{et~al.}(2023)Ebbesen, Rubio, and
  Scholes]{Ebbesen2023-fd}
Ebbesen,~T.~W.; Rubio,~A.; Scholes,~G.~D. Introduction: Polaritonic Chemistry.
  \emph{Chem. Rev.} \textbf{2023}, \emph{123}, 12037--12038\relax
\mciteBstWouldAddEndPuncttrue
\mciteSetBstMidEndSepPunct{\mcitedefaultmidpunct}
{\mcitedefaultendpunct}{\mcitedefaultseppunct}\relax
\EndOfBibitem
\bibitem[Bhuyan \latin{et~al.}(2023)Bhuyan, Mony, Kotov, Castellanos,
  G{\'o}mez~Rivas, Shegai, and B{\"o}rjesson]{Bhuyan2023-se}
Bhuyan,~R.; Mony,~J.; Kotov,~O.; Castellanos,~G.~W.; G{\'o}mez~Rivas,~J.;
  Shegai,~T.~O.; B{\"o}rjesson,~K. The Rise and Current Status of Polaritonic
  Photochemistry and Photophysics. \emph{Chem. Rev.} \textbf{2023}, \emph{123},
  10877--10919\relax
\mciteBstWouldAddEndPuncttrue
\mciteSetBstMidEndSepPunct{\mcitedefaultmidpunct}
{\mcitedefaultendpunct}{\mcitedefaultseppunct}\relax
\EndOfBibitem
\bibitem[Dovzhenko \latin{et~al.}(2018)Dovzhenko, Ryabchuk, Rakovich, and
  Nabiev]{Dovzhenko2018-ph}
Dovzhenko,~D.~S.; Ryabchuk,~S.~V.; Rakovich,~Y.~P.; Nabiev,~I.~R. Light-matter
  interaction in the strong coupling regime: configurations, conditions, and
  applications. \emph{Nanoscale} \textbf{2018}, \emph{10}, 3589--3605\relax
\mciteBstWouldAddEndPuncttrue
\mciteSetBstMidEndSepPunct{\mcitedefaultmidpunct}
{\mcitedefaultendpunct}{\mcitedefaultseppunct}\relax
\EndOfBibitem
\bibitem[Hertzog \latin{et~al.}(2019)Hertzog, Wang, Mony, and
  Börjesson]{Hertzog2019-uh}
Hertzog,~M.; Wang,~M.; Mony,~J.; Börjesson,~K. Strong light-matter
  interactions: a new direction within chemistry. \emph{Chem. Soc. Rev.}
  \textbf{2019}, \emph{48}, 937--961\relax
\mciteBstWouldAddEndPuncttrue
\mciteSetBstMidEndSepPunct{\mcitedefaultmidpunct}
{\mcitedefaultendpunct}{\mcitedefaultseppunct}\relax
\EndOfBibitem
\bibitem[Thomas \latin{et~al.}(2016)Thomas, George, Shalabney, Dryzhakov,
  Varma, Moran, Chervy, Zhong, Devaux, Genet, Hutchison, and
  Ebbesen]{Thomas2016-fy}
Thomas,~A.; George,~J.; Shalabney,~A.; Dryzhakov,~M.; Varma,~S.~J.; Moran,~J.;
  Chervy,~T.; Zhong,~X.; Devaux,~E.; Genet,~C.; Hutchison,~J.~A.;
  Ebbesen,~T.~W. Ground-State Chemical Reactivity under Vibrational Coupling to
  the Vacuum Electromagnetic Field. \emph{Angew. Chem. Int. Ed Engl.}
  \textbf{2016}, \emph{55}, 11462--11466\relax
\mciteBstWouldAddEndPuncttrue
\mciteSetBstMidEndSepPunct{\mcitedefaultmidpunct}
{\mcitedefaultendpunct}{\mcitedefaultseppunct}\relax
\EndOfBibitem
\bibitem[Lather \latin{et~al.}(2019)Lather, Bhatt, Thomas, Ebbesen, and
  George]{Lather2019-yn}
Lather,~J.; Bhatt,~P.; Thomas,~A.; Ebbesen,~T.~W.; George,~J. Cavity Catalysis
  by Cooperative Vibrational Strong Coupling of Reactant and Solvent Molecules.
  \emph{Angew. Chem. Int. Ed Engl.} \textbf{2019}, \emph{58},
  10635--10638\relax
\mciteBstWouldAddEndPuncttrue
\mciteSetBstMidEndSepPunct{\mcitedefaultmidpunct}
{\mcitedefaultendpunct}{\mcitedefaultseppunct}\relax
\EndOfBibitem
\bibitem[Hirai \latin{et~al.}(2020)Hirai, Takeda, Hutchison, and
  Uji-I]{Hirai2020-uv}
Hirai,~K.; Takeda,~R.; Hutchison,~J.~A.; Uji-I,~H. Modulation of Prins
  Cyclization by Vibrational Strong Coupling. \emph{Angew. Chem. Int. Ed Engl.}
  \textbf{2020}, \emph{59}, 5332--5335\relax
\mciteBstWouldAddEndPuncttrue
\mciteSetBstMidEndSepPunct{\mcitedefaultmidpunct}
{\mcitedefaultendpunct}{\mcitedefaultseppunct}\relax
\EndOfBibitem
\bibitem[Ahn \latin{et~al.}(2023)Ahn, Triana, Recabal, Herrera, and
  Simpkins]{Ahn2023-qk}
Ahn,~W.; Triana,~J.~F.; Recabal,~F.; Herrera,~F.; Simpkins,~B.~S. Modification
  of ground-state chemical reactivity via light-matter coherence in infrared
  cavities. \emph{Science} \textbf{2023}, \emph{380}, 1165--1168\relax
\mciteBstWouldAddEndPuncttrue
\mciteSetBstMidEndSepPunct{\mcitedefaultmidpunct}
{\mcitedefaultendpunct}{\mcitedefaultseppunct}\relax
\EndOfBibitem
\bibitem[Patrahau \latin{et~al.}(2024)Patrahau, Piejko, Mayer, Antheaume,
  Sangchai, Ragazzon, Jayachandran, Devaux, Genet, Moran, and
  Ebbesen]{Patrahau2024-zx}
Patrahau,~B.; Piejko,~M.; Mayer,~R.~J.; Antheaume,~C.; Sangchai,~T.;
  Ragazzon,~G.; Jayachandran,~A.; Devaux,~E.; Genet,~C.; Moran,~J.;
  Ebbesen,~T.~W. Direct observation of polaritonic chemistry by nuclear
  magnetic resonance spectroscopy. \emph{Angew. Chem. Weinheim Bergstr. Ger.}
  \textbf{2024}, \emph{136}, e202401368\relax
\mciteBstWouldAddEndPuncttrue
\mciteSetBstMidEndSepPunct{\mcitedefaultmidpunct}
{\mcitedefaultendpunct}{\mcitedefaultseppunct}\relax
\EndOfBibitem
\bibitem[Brawley \latin{et~al.}(2025)Brawley, Pannir-Sivajothi, Yim, Poh,
  Yuen-Zhou, and Sheldon]{Brawley2025-qh}
Brawley,~Z.~T.; Pannir-Sivajothi,~S.; Yim,~J.~E.; Poh,~Y.~R.; Yuen-Zhou,~J.;
  Sheldon,~M. Vibrational weak and strong coupling modify a chemical reaction
  via cavity-mediated radiative energy transfer. \emph{Nat. Chem.}
  \textbf{2025}, 1--9\relax
\mciteBstWouldAddEndPuncttrue
\mciteSetBstMidEndSepPunct{\mcitedefaultmidpunct}
{\mcitedefaultendpunct}{\mcitedefaultseppunct}\relax
\EndOfBibitem
\bibitem[Sch{\"u}tz \latin{et~al.}(2020)Sch{\"u}tz, Schachenmayer,
  Hagenm{\"u}ller, Brennen, Volz, Sandoghdar, Ebbesen, Genes, and
  Pupillo]{Schutz2020-en}
Sch{\"u}tz,~S.; Schachenmayer,~J.; Hagenm{\"u}ller,~D.; Brennen,~G.~K.;
  Volz,~T.; Sandoghdar,~V.; Ebbesen,~T.~W.; Genes,~C.; Pupillo,~G.
  {Ensemble-Induced} Strong {Light-Matter} Coupling of a Single Quantum
  Emitter. \emph{Phys. Rev. Lett.} \textbf{2020}, \emph{124}, 113602\relax
\mciteBstWouldAddEndPuncttrue
\mciteSetBstMidEndSepPunct{\mcitedefaultmidpunct}
{\mcitedefaultendpunct}{\mcitedefaultseppunct}\relax
\EndOfBibitem
\bibitem[Sidler \latin{et~al.}(2022)Sidler, Ruggenthaler, Sch{\"a}fer, Ronca,
  and Rubio]{Sidler2022-cg}
Sidler,~D.; Ruggenthaler,~M.; Sch{\"a}fer,~C.; Ronca,~E.; Rubio,~A. A
  perspective on ab initio modeling of polaritonic chemistry: The role of
  non-equilibrium effects and quantum collectivity. \emph{J. Chem. Phys.}
  \textbf{2022}, \emph{156}, 230901\relax
\mciteBstWouldAddEndPuncttrue
\mciteSetBstMidEndSepPunct{\mcitedefaultmidpunct}
{\mcitedefaultendpunct}{\mcitedefaultseppunct}\relax
\EndOfBibitem
\bibitem[Taylor \latin{et~al.}(2022)Taylor, Mandal, and Huo]{Taylor2022-pv}
Taylor,~M. A.~D.; Mandal,~A.; Huo,~P. Resolving ambiguities of the mode
  truncation in cavity quantum electrodynamics. \emph{Opt. Lett.}
  \textbf{2022}, \emph{47}, 1446--1449\relax
\mciteBstWouldAddEndPuncttrue
\mciteSetBstMidEndSepPunct{\mcitedefaultmidpunct}
{\mcitedefaultendpunct}{\mcitedefaultseppunct}\relax
\EndOfBibitem
\bibitem[Sánchez-Barquilla \latin{et~al.}(2022)Sánchez-Barquilla,
  Fernández-Domínguez, Feist, and García-Vidal]{Sanchez-Barquilla2022-dq}
Sánchez-Barquilla,~M.; Fernández-Domínguez,~A.~I.; Feist,~J.;
  García-Vidal,~F.~J. A Theoretical Perspective on Molecular Polaritonics.
  \emph{ACS Photonics} \textbf{2022}, \relax
\mciteBstWouldAddEndPunctfalse
\mciteSetBstMidEndSepPunct{\mcitedefaultmidpunct}
{}{\mcitedefaultseppunct}\relax
\EndOfBibitem
\bibitem[Simpkins \latin{et~al.}(2023)Simpkins, Dunkelberger, and
  Vurgaftman]{Simpkins2023-ze}
Simpkins,~B.~S.; Dunkelberger,~A.~D.; Vurgaftman,~I. Control, Modulation, and
  Analytical Descriptions of Vibrational Strong Coupling. \emph{Chem. Rev.}
  \textbf{2023}, \relax
\mciteBstWouldAddEndPunctfalse
\mciteSetBstMidEndSepPunct{\mcitedefaultmidpunct}
{}{\mcitedefaultseppunct}\relax
\EndOfBibitem
\bibitem[Schnappinger \latin{et~al.}(2023)Schnappinger, Sidler, Ruggenthaler,
  Rubio, and Kowalewski]{Schnappinger2023-hh}
Schnappinger,~T.; Sidler,~D.; Ruggenthaler,~M.; Rubio,~A.; Kowalewski,~M.
  Cavity {Born-Oppenheimer} {Hartree-Fock} Ansatz: {Light-Matter} Properties of
  Strongly Coupled Molecular Ensembles. \emph{J. Phys. Chem. Lett.}
  \textbf{2023}, \emph{14}, 8024--8033\relax
\mciteBstWouldAddEndPuncttrue
\mciteSetBstMidEndSepPunct{\mcitedefaultmidpunct}
{\mcitedefaultendpunct}{\mcitedefaultseppunct}\relax
\EndOfBibitem
\bibitem[Svendsen \latin{et~al.}(2023)Svendsen, Ruggenthaler, Hübener,
  Schäfer, Eckstein, Rubio, and Latini]{Svendsen2023-ly}
Svendsen,~M.~K.; Ruggenthaler,~M.; Hübener,~H.; Schäfer,~C.; Eckstein,~M.;
  Rubio,~A.; Latini,~S. Theory of quantum light-matter interaction in cavities:
  Extended systems and the long wavelength approximation. \emph{arXiv
  [cond-mat.mes-hall]} \textbf{2023}, \relax
\mciteBstWouldAddEndPunctfalse
\mciteSetBstMidEndSepPunct{\mcitedefaultmidpunct}
{}{\mcitedefaultseppunct}\relax
\EndOfBibitem
\bibitem[Sidler \latin{et~al.}(2024)Sidler, Schnappinger, Obzhirov,
  Ruggenthaler, Kowalewski, and Rubio]{Sidler2024-tw}
Sidler,~D.; Schnappinger,~T.; Obzhirov,~A.; Ruggenthaler,~M.; Kowalewski,~M.;
  Rubio,~A. Unraveling a cavity-induced molecular polarization mechanism from
  collective vibrational strong coupling. \emph{J. Phys. Chem. Lett.}
  \textbf{2024}, \emph{15}, 5208--5214\relax
\mciteBstWouldAddEndPuncttrue
\mciteSetBstMidEndSepPunct{\mcitedefaultmidpunct}
{\mcitedefaultendpunct}{\mcitedefaultseppunct}\relax
\EndOfBibitem
\bibitem[Michon and Simpkins(2024)Michon, and Simpkins]{Michon2024-vb}
Michon,~M.~A.; Simpkins,~B.~S. Impact of cavity length non-uniformity on
  reaction rate extraction in strong coupling experiments. \emph{J. Am. Chem.
  Soc.} \textbf{2024}, \relax
\mciteBstWouldAddEndPunctfalse
\mciteSetBstMidEndSepPunct{\mcitedefaultmidpunct}
{}{\mcitedefaultseppunct}\relax
\EndOfBibitem
\bibitem[Nelson and Weichman(2024)Nelson, and Weichman]{Nelson2024-ym}
Nelson,~J.~C.; Weichman,~M.~L. More than just smoke and mirrors: Gas-phase
  polaritons for optical control of chemistry. \emph{J. Chem. Phys.}
  \textbf{2024}, \emph{161}, 074304\relax
\mciteBstWouldAddEndPuncttrue
\mciteSetBstMidEndSepPunct{\mcitedefaultmidpunct}
{\mcitedefaultendpunct}{\mcitedefaultseppunct}\relax
\EndOfBibitem
\bibitem[Ying \latin{et~al.}(2024)Ying, Taylor, and Huo]{Ying2024-aj}
Ying,~W.; Taylor,~M. A.~D.; Huo,~P. Resonance theory of vibrational polariton
  chemistry at the normal incidence. \emph{Nanophotonics} \textbf{2024}, \relax
\mciteBstWouldAddEndPunctfalse
\mciteSetBstMidEndSepPunct{\mcitedefaultmidpunct}
{}{\mcitedefaultseppunct}\relax
\EndOfBibitem
\bibitem[Sánchez~Martínez \latin{et~al.}(2024)Sánchez~Martínez, Feist, and
  García-Vidal]{Sanchez_Martinez2024-wu}
Sánchez~Martínez,~C.~J.; Feist,~J.; García-Vidal,~F.~J. A mixed
  perturbative-nonperturbative treatment for strong light-matter interactions.
  \emph{Nanophotonics} \textbf{2024}, \relax
\mciteBstWouldAddEndPunctfalse
\mciteSetBstMidEndSepPunct{\mcitedefaultmidpunct}
{}{\mcitedefaultseppunct}\relax
\EndOfBibitem
\bibitem[Vasil \latin{et~al.}(2025)Vasil, Ilia, and H.]{Vasil2025-by}
Vasil,~R.; Ilia,~T.; H.,~R.~S. Cavity-mediated collective resonant suppression
  of local molecular vibrations. \emph{arXiv [quant-ph]} \textbf{2025}, \relax
\mciteBstWouldAddEndPunctfalse
\mciteSetBstMidEndSepPunct{\mcitedefaultmidpunct}
{}{\mcitedefaultseppunct}\relax
\EndOfBibitem
\bibitem[Haugland \latin{et~al.}(2020)Haugland, Ronca, Kjønstad, Rubio, and
  Koch]{Haugland2020-xh}
Haugland,~T.~S.; Ronca,~E.; Kjønstad,~E.~F.; Rubio,~A.; Koch,~H. Coupled
  Cluster Theory for Molecular Polaritons: Changing Ground and Excited States.
  \emph{Phys. Rev. X} \textbf{2020}, \emph{10}, 041043\relax
\mciteBstWouldAddEndPuncttrue
\mciteSetBstMidEndSepPunct{\mcitedefaultmidpunct}
{\mcitedefaultendpunct}{\mcitedefaultseppunct}\relax
\EndOfBibitem
\bibitem[Angelico \latin{et~al.}(2023)Angelico, Haugland, Ronca, and
  Koch]{Angelico2023-jh}
Angelico,~S.; Haugland,~T.~S.; Ronca,~E.; Koch,~H. Coupled cluster cavity
  Born-Oppenheimer approximation for electronic strong coupling. \emph{J. Chem.
  Phys.} \textbf{2023}, \emph{159}, 214112\relax
\mciteBstWouldAddEndPuncttrue
\mciteSetBstMidEndSepPunct{\mcitedefaultmidpunct}
{\mcitedefaultendpunct}{\mcitedefaultseppunct}\relax
\EndOfBibitem
\bibitem[Schnappinger and Kowalewski(2023)Schnappinger, and
  Kowalewski]{Schnappinger2023-wp}
Schnappinger,~T.; Kowalewski,~M. Ab Initio Vibro-Polaritonic Spectra in
  Strongly Coupled Cavity-Molecule Systems. \emph{J. Chem. Theory Comput.}
  \textbf{2023}, \emph{19}, 9278--9289\relax
\mciteBstWouldAddEndPuncttrue
\mciteSetBstMidEndSepPunct{\mcitedefaultmidpunct}
{\mcitedefaultendpunct}{\mcitedefaultseppunct}\relax
\EndOfBibitem
\bibitem[Fischer(2024)]{Fischer2024-jb}
Fischer,~E.~W. Cavity-modified local and non-local electronic interactions in
  molecular ensembles under vibrational strong coupling. \emph{J. Chem. Phys.}
  \textbf{2024}, \emph{161}, 164112\relax
\mciteBstWouldAddEndPuncttrue
\mciteSetBstMidEndSepPunct{\mcitedefaultmidpunct}
{\mcitedefaultendpunct}{\mcitedefaultseppunct}\relax
\EndOfBibitem
\bibitem[Castagnola \latin{et~al.}(2024)Castagnola, Haugland, Ronca, Koch, and
  Schäfer]{Castagnola2024-vt}
Castagnola,~M.; Haugland,~T.~S.; Ronca,~E.; Koch,~H.; Schäfer,~C. Collective
  Strong Coupling Modifies Aggregation and Solvation. \emph{J. Phys. Chem.
  Lett.} \textbf{2024}, 1428--1434\relax
\mciteBstWouldAddEndPuncttrue
\mciteSetBstMidEndSepPunct{\mcitedefaultmidpunct}
{\mcitedefaultendpunct}{\mcitedefaultseppunct}\relax
\EndOfBibitem
\bibitem[Flick \latin{et~al.}(2017)Flick, Ruggenthaler, Appel, and
  Rubio]{flick2017atoms}
Flick,~J.; Ruggenthaler,~M.; Appel,~H.; Rubio,~A. Atoms and molecules in
  cavities, from weak to strong coupling in quantum-electrodynamics (QED)
  chemistry. \emph{Proc. Natl. Acad. Sci. U.S.A.} \textbf{2017}, \emph{114},
  3026--3034\relax
\mciteBstWouldAddEndPuncttrue
\mciteSetBstMidEndSepPunct{\mcitedefaultmidpunct}
{\mcitedefaultendpunct}{\mcitedefaultseppunct}\relax
\EndOfBibitem
\bibitem[Flick \latin{et~al.}(2017)Flick, Appel, Ruggenthaler, and
  Rubio]{flick2017cavity}
Flick,~J.; Appel,~H.; Ruggenthaler,~M.; Rubio,~A. Cavity Born--Oppenheimer
  approximation for correlated electron--nuclear-photon systems. \emph{J. Chem.
  Theory Comput.} \textbf{2017}, \emph{13}, 1616--1625\relax
\mciteBstWouldAddEndPuncttrue
\mciteSetBstMidEndSepPunct{\mcitedefaultmidpunct}
{\mcitedefaultendpunct}{\mcitedefaultseppunct}\relax
\EndOfBibitem
\bibitem[Flick and Narang(2018)Flick, and Narang]{Flick2018-ns}
Flick,~J.; Narang,~P. {Cavity-Correlated} {Electron-Nuclear} Dynamics from
  First Principles. \emph{Phys. Rev. Lett.} \textbf{2018}, \emph{121},
  113002\relax
\mciteBstWouldAddEndPuncttrue
\mciteSetBstMidEndSepPunct{\mcitedefaultmidpunct}
{\mcitedefaultendpunct}{\mcitedefaultseppunct}\relax
\EndOfBibitem
\bibitem[Bernath(2005)]{Bernath2005-bt}
Bernath,~P.~F. \emph{Spectra of atoms and molecules}, 2nd ed.; Oxford
  University Press: New York, NY, 2005\relax
\mciteBstWouldAddEndPuncttrue
\mciteSetBstMidEndSepPunct{\mcitedefaultmidpunct}
{\mcitedefaultendpunct}{\mcitedefaultseppunct}\relax
\EndOfBibitem
\bibitem[London(1937)]{London1937-um}
London,~F. The general theory of molecular forces. \emph{Trans. Faraday Soc.}
  \textbf{1937}, \emph{33}, 8b\relax
\mciteBstWouldAddEndPuncttrue
\mciteSetBstMidEndSepPunct{\mcitedefaultmidpunct}
{\mcitedefaultendpunct}{\mcitedefaultseppunct}\relax
\EndOfBibitem
\bibitem[Maitland \latin{et~al.}(1983)Maitland, Rigby, Smith, Wakeham, and
  Henderson]{Maitland1983IntermolecularFT}
Maitland,~G.; Rigby,~M.; Smith,~E.~B.; Wakeham,~W.~A.; Henderson,~D.~A.
  Intermolecular Forces: Their Origin and Determination. \emph{Physics Today}
  \textbf{1983}, \emph{36}, 57--58\relax
\mciteBstWouldAddEndPuncttrue
\mciteSetBstMidEndSepPunct{\mcitedefaultmidpunct}
{\mcitedefaultendpunct}{\mcitedefaultseppunct}\relax
\EndOfBibitem
\bibitem[Hait and Head-Gordon(2023)Hait, and Head-Gordon]{Hait2023-do}
Hait,~D.; Head-Gordon,~M. When is a bond broken? The polarizability
  perspective. \emph{Angew. Chem. Int. Ed Engl.} \textbf{2023}, \emph{62},
  e202312078\relax
\mciteBstWouldAddEndPuncttrue
\mciteSetBstMidEndSepPunct{\mcitedefaultmidpunct}
{\mcitedefaultendpunct}{\mcitedefaultseppunct}\relax
\EndOfBibitem
\bibitem[Schnappinger and Kowalewski(2024)Schnappinger, and
  Kowalewski]{Schnappinger2024-vt}
Schnappinger,~T.; Kowalewski,~M. Do Molecular Geometries Change Under
  Vibrational Strong Coupling? \emph{J. Phys. Chem. Lett.} \textbf{2024},
  \emph{15}, 7700--7707\relax
\mciteBstWouldAddEndPuncttrue
\mciteSetBstMidEndSepPunct{\mcitedefaultmidpunct}
{\mcitedefaultendpunct}{\mcitedefaultseppunct}\relax
\EndOfBibitem
\bibitem[Horak \latin{et~al.}(2024)Horak, Sidler, Schnappinger, Huang,
  Ruggenthaler, and Rubio]{Horak2024-fa}
Horak,~J.; Sidler,~D.; Schnappinger,~T.; Huang,~W.-M.; Ruggenthaler,~M.;
  Rubio,~A. Analytic model reveals local molecular polarizability changes
  induced by collective strong coupling in optical cavities. \emph{Phys. Rev.
  Res.} \textbf{2024}, \emph{7}, 013242\relax
\mciteBstWouldAddEndPuncttrue
\mciteSetBstMidEndSepPunct{\mcitedefaultmidpunct}
{\mcitedefaultendpunct}{\mcitedefaultseppunct}\relax
\EndOfBibitem
\bibitem[Shalabney \latin{et~al.}(2015)Shalabney, George, Hiura, Hutchison,
  Genet, Hellwig, and Ebbesen]{Shalabney2015-qs}
Shalabney,~A.; George,~J.; Hiura,~H.; Hutchison,~J.~A.; Genet,~C.; Hellwig,~P.;
  Ebbesen,~T.~W. Enhanced Raman Scattering from Vibro-Polariton Hybrid States.
  \emph{Angew. Chem. Int. Ed Engl.} \textbf{2015}, \emph{54}, 7971--7975\relax
\mciteBstWouldAddEndPuncttrue
\mciteSetBstMidEndSepPunct{\mcitedefaultmidpunct}
{\mcitedefaultendpunct}{\mcitedefaultseppunct}\relax
\EndOfBibitem
\bibitem[Takele \latin{et~al.}(2021)Takele, Piatkowski, Wackenhut, Gawinkowski,
  Meixner, and Waluk]{Takele2021-xv}
Takele,~W.~M.; Piatkowski,~L.; Wackenhut,~F.; Gawinkowski,~S.; Meixner,~A.~J.;
  Waluk,~J. Scouting for strong light-matter coupling signatures in Raman
  spectra. \emph{Phys. Chem. Chem. Phys.} \textbf{2021}, \emph{23},
  16837--16846\relax
\mciteBstWouldAddEndPuncttrue
\mciteSetBstMidEndSepPunct{\mcitedefaultmidpunct}
{\mcitedefaultendpunct}{\mcitedefaultseppunct}\relax
\EndOfBibitem
\bibitem[Verdelli \latin{et~al.}(2022)Verdelli, Schulpen, Baldi, and
  Rivas]{Verdelli2022-yl}
Verdelli,~F.; Schulpen,~J. J. P.~M.; Baldi,~A.; Rivas,~J.~G. Chasing
  vibro-polariton fingerprints in infrared and Raman spectra using surface
  lattice resonances on extended metasurfaces. \emph{J. Phys. Chem. C
  Nanomater. Interfaces} \textbf{2022}, \emph{126}, 7143--7151\relax
\mciteBstWouldAddEndPuncttrue
\mciteSetBstMidEndSepPunct{\mcitedefaultmidpunct}
{\mcitedefaultendpunct}{\mcitedefaultseppunct}\relax
\EndOfBibitem
\bibitem[del Pino \latin{et~al.}(2015)del Pino, Feist, and
  Garcia-Vidal]{del-Pino2015-un}
del Pino,~J.; Feist,~J.; Garcia-Vidal,~F.~J. Signatures of vibrational strong
  coupling in Raman scattering. \emph{J. Phys. Chem. C Nanomater. Interfaces}
  \textbf{2015}, \emph{119}, 29132--29137\relax
\mciteBstWouldAddEndPuncttrue
\mciteSetBstMidEndSepPunct{\mcitedefaultmidpunct}
{\mcitedefaultendpunct}{\mcitedefaultseppunct}\relax
\EndOfBibitem
\bibitem[Strashko and Keeling(2016)Strashko, and Keeling]{Strashko2016-nt}
Strashko,~A.; Keeling,~J. Raman scattering with strongly coupled
  vibron-polaritons. \emph{Phys. Rev. A} \textbf{2016}, \emph{94}, 023843\relax
\mciteBstWouldAddEndPuncttrue
\mciteSetBstMidEndSepPunct{\mcitedefaultmidpunct}
{\mcitedefaultendpunct}{\mcitedefaultseppunct}\relax
\EndOfBibitem
\bibitem[Huang and Liang(2024)Huang, and Liang]{Huang2024-is}
Huang,~X.; Liang,~W. Analytical Derivative Approaches for Vibro-Polaritionic
  Structures and Properties. \emph{ChemRxiv} \textbf{2024}, \relax
\mciteBstWouldAddEndPunctfalse
\mciteSetBstMidEndSepPunct{\mcitedefaultmidpunct}
{}{\mcitedefaultseppunct}\relax
\EndOfBibitem
\bibitem[Huang and Liang(2025)Huang, and Liang]{Huang2025-lu}
Huang,~X.; Liang,~W. Analytical derivative approaches for vibro-polaritonic
  structures and properties. {I}. Formalism and implementation. \emph{J. Chem.
  Phys.} \textbf{2025}, \emph{162}, 024115\relax
\mciteBstWouldAddEndPuncttrue
\mciteSetBstMidEndSepPunct{\mcitedefaultmidpunct}
{\mcitedefaultendpunct}{\mcitedefaultseppunct}\relax
\EndOfBibitem
\bibitem[Sch{\"a}fer \latin{et~al.}(2018)Sch{\"a}fer, Ruggenthaler, and
  Rubio]{Schafer2018-vf}
Sch{\"a}fer,~C.; Ruggenthaler,~M.; Rubio,~A. Ab initio nonrelativistic quantum
  electrodynamics: Bridging quantum chemistry and quantum optics from weak to
  strong coupling. \emph{Phys. Rev. A} \textbf{2018}, \emph{98}, 043801\relax
\mciteBstWouldAddEndPuncttrue
\mciteSetBstMidEndSepPunct{\mcitedefaultmidpunct}
{\mcitedefaultendpunct}{\mcitedefaultseppunct}\relax
\EndOfBibitem
\bibitem[Ruggenthaler \latin{et~al.}(2023)Ruggenthaler, Sidler, and
  Rubio]{Ruggenthaler2023-aa}
Ruggenthaler,~M.; Sidler,~D.; Rubio,~A. Understanding Polaritonic Chemistry
  from Ab Initio Quantum Electrodynamics. \emph{Chem. Rev.} \textbf{2023},
  \relax
\mciteBstWouldAddEndPunctfalse
\mciteSetBstMidEndSepPunct{\mcitedefaultmidpunct}
{}{\mcitedefaultseppunct}\relax
\EndOfBibitem
\bibitem[Rokaj \latin{et~al.}(2018)Rokaj, Welakuh, Ruggenthaler, and
  Rubio]{Rokaj2018-ww}
Rokaj,~V.; Welakuh,~D.~M.; Ruggenthaler,~M.; Rubio,~A. Light--matter
  interaction in the long-wavelength limit: no ground-state without dipole
  self-energy. \emph{J. Phys. B At. Mol. Opt. Phys.} \textbf{2018}, \emph{51},
  034005\relax
\mciteBstWouldAddEndPuncttrue
\mciteSetBstMidEndSepPunct{\mcitedefaultmidpunct}
{\mcitedefaultendpunct}{\mcitedefaultseppunct}\relax
\EndOfBibitem
\bibitem[Sch{\"a}fer \latin{et~al.}(2020)Sch{\"a}fer, Ruggenthaler, Rokaj, and
  Rubio]{Schafer2020-cb}
Sch{\"a}fer,~C.; Ruggenthaler,~M.; Rokaj,~V.; Rubio,~A. Relevance of the
  Quadratic Diamagnetic and {Self-Polarization} Terms in Cavity Quantum
  Electrodynamics. \emph{ACS Photonics} \textbf{2020}, \emph{7}, 975--990\relax
\mciteBstWouldAddEndPuncttrue
\mciteSetBstMidEndSepPunct{\mcitedefaultmidpunct}
{\mcitedefaultendpunct}{\mcitedefaultseppunct}\relax
\EndOfBibitem
\bibitem[Szabo and Ostlund(1996)Szabo, and Ostlund]{aszabo82-qc}
Szabo,~A.; Ostlund,~N.~S. \emph{Modern Quantum Chemistry: Introduction to
  Advanced Electronic Structure Theory}, 1st ed.; Dover Publications, Inc.:
  Mineola, 1996\relax
\mciteBstWouldAddEndPuncttrue
\mciteSetBstMidEndSepPunct{\mcitedefaultmidpunct}
{\mcitedefaultendpunct}{\mcitedefaultseppunct}\relax
\EndOfBibitem
\bibitem[Atkins and Friedman(2011)Atkins, and Friedman]{atkins_molecular_2011}
Atkins,~P.~W.; Friedman,~R. \emph{Molecular quantum mechanics}, 5th ed.; Oxford
  Univ. Press: Oxford, 2011\relax
\mciteBstWouldAddEndPuncttrue
\mciteSetBstMidEndSepPunct{\mcitedefaultmidpunct}
{\mcitedefaultendpunct}{\mcitedefaultseppunct}\relax
\EndOfBibitem
\bibitem[Brakestad \latin{et~al.}(2020)Brakestad, Jensen, Wind, D'Alessandro,
  Genovese, Hopmann, and Frediani]{Brakestad2020-bo}
Brakestad,~A.; Jensen,~S.~R.; Wind,~P.; D'Alessandro,~M.; Genovese,~L.;
  Hopmann,~K.~H.; Frediani,~L. Static polarizabilities at the basis set limit:
  A benchmark of 124 species. \emph{J. Chem. Theory Comput.} \textbf{2020},
  \emph{16}, 4874--4882\relax
\mciteBstWouldAddEndPuncttrue
\mciteSetBstMidEndSepPunct{\mcitedefaultmidpunct}
{\mcitedefaultendpunct}{\mcitedefaultseppunct}\relax
\EndOfBibitem
\bibitem[Helgaker and Jørgensen(1988)Helgaker, and
  Jørgensen]{Helgaker1988-if}
Helgaker,~T.; Jørgensen,~P. \emph{Advances in Quantum Chemistry}; Advances in
  quantum chemistry; Elsevier, 1988; Vol.~19; pp 183--245\relax
\mciteBstWouldAddEndPuncttrue
\mciteSetBstMidEndSepPunct{\mcitedefaultmidpunct}
{\mcitedefaultendpunct}{\mcitedefaultseppunct}\relax
\EndOfBibitem
\bibitem[Neugebauer \latin{et~al.}(2002)Neugebauer, Reiher, Kind, and
  Hess]{Neugebauer2002-nh}
Neugebauer,~J.; Reiher,~M.; Kind,~C.; Hess,~B.~A. Quantum chemical calculation
  of vibrational spectra of large molecules--Raman and {IR} spectra for
  Buckminsterfullerene. \emph{J. Comput. Chem.} \textbf{2002}, \emph{23},
  895--910\relax
\mciteBstWouldAddEndPuncttrue
\mciteSetBstMidEndSepPunct{\mcitedefaultmidpunct}
{\mcitedefaultendpunct}{\mcitedefaultseppunct}\relax
\EndOfBibitem
\bibitem[Smith \latin{et~al.}(2018)Smith, Burns, Sirianni, Nascimento, Kumar,
  James, Schriber, Zhang, Zhang, Abbott, Berquist, Lechner, Cunha, Heide,
  Waldrop, Takeshita, Alenaizan, Neuhauser, King, Simmonett, Turney, Schaefer,
  Evangelista, DePrince, Crawford, Patkowski, and Sherrill]{Smith2018-tu}
Smith,~D. G.~A.; Burns,~L.~A.; Sirianni,~D.~A.; Nascimento,~D.~R.; Kumar,~A.;
  James,~A.~M.; Schriber,~J.~B.; Zhang,~T.; Zhang,~B.; Abbott,~A.~S.;
  Berquist,~E.~J.; Lechner,~M.~H.; Cunha,~L.~A.; Heide,~A.~G.; Waldrop,~J.~M.;
  Takeshita,~T.~Y.; Alenaizan,~A.; Neuhauser,~D.; King,~R.~A.;
  Simmonett,~A.~C.; Turney,~J.~M.; Schaefer,~H.~F.; Evangelista,~F.~A.;
  DePrince,~A.~E.,~3rd; Crawford,~T.~D.; Patkowski,~K.; Sherrill,~C.~D.
  {Psi4NumPy}: An Interactive Quantum Chemistry Programming Environment for
  Reference Implementations and Rapid Development. \emph{J. Chem. Theory
  Comput.} \textbf{2018}, \emph{14}, 3504--3511\relax
\mciteBstWouldAddEndPuncttrue
\mciteSetBstMidEndSepPunct{\mcitedefaultmidpunct}
{\mcitedefaultendpunct}{\mcitedefaultseppunct}\relax
\EndOfBibitem
\bibitem[Smith \latin{et~al.}(2020)Smith, Burns, Simmonett, Parrish, Schieber,
  Galvelis, Kraus, Kruse, Di~Remigio, Alenaizan, James, Lehtola, Misiewicz,
  Scheurer, Shaw, Schriber, Xie, Glick, Sirianni, O'Brien, Waldrop, Kumar,
  Hohenstein, Pritchard, Brooks, Schaefer, Sokolov, Patkowski, DePrince,
  Bozkaya, King, Evangelista, Turney, Crawford, and Sherrill]{Smith2020-kq}
Smith,~D. G.~A.; Burns,~L.~A.; Simmonett,~A.~C.; Parrish,~R.~M.;
  Schieber,~M.~C.; Galvelis,~R.; Kraus,~P.; Kruse,~H.; Di~Remigio,~R.;
  Alenaizan,~A.; James,~A.~M.; Lehtola,~S.; Misiewicz,~J.~P.; Scheurer,~M.;
  Shaw,~R.~A.; Schriber,~J.~B.; Xie,~Y.; Glick,~Z.~L.; Sirianni,~D.~A.;
  O'Brien,~J.~S.; Waldrop,~J.~M.; Kumar,~A.; Hohenstein,~E.~G.;
  Pritchard,~B.~P.; Brooks,~B.~R.; Schaefer,~H.~F.,~3rd; Sokolov,~A.~Y.;
  Patkowski,~K.; DePrince,~A.~E.,~3rd; Bozkaya,~U.; King,~R.~A.;
  Evangelista,~F.~A.; Turney,~J.~M.; Crawford,~T.~D.; Sherrill,~C.~D. Psi4 1.4:
  Open-source software for high-throughput quantum chemistry. \emph{J. Chem.
  Phys.} \textbf{2020}, \emph{152}, 184108\relax
\mciteBstWouldAddEndPuncttrue
\mciteSetBstMidEndSepPunct{\mcitedefaultmidpunct}
{\mcitedefaultendpunct}{\mcitedefaultseppunct}\relax
\EndOfBibitem
\bibitem[Kendall \latin{et~al.}(1992)Kendall, Dunning, and
  Harrison]{Kendall1992-wu}
Kendall,~R.~A.; Dunning,~T.~H.; Harrison,~R.~J. Electron affinities of the
  first‐row atoms revisited. Systematic basis sets and wave functions.
  \emph{J. Chem. Phys.} \textbf{1992}, \emph{96}, 6796--6806\relax
\mciteBstWouldAddEndPuncttrue
\mciteSetBstMidEndSepPunct{\mcitedefaultmidpunct}
{\mcitedefaultendpunct}{\mcitedefaultseppunct}\relax
\EndOfBibitem
\bibitem[Liebenthal and DePrince(2024)Liebenthal, and
  DePrince]{Liebenthal2024-mx}
Liebenthal,~M.~D.; DePrince,~A.~E.,~3rd The orientation dependence of
  cavity-modified chemistry. \emph{J. Chem. Phys.} \textbf{2024},
  \emph{161}\relax
\mciteBstWouldAddEndPuncttrue
\mciteSetBstMidEndSepPunct{\mcitedefaultmidpunct}
{\mcitedefaultendpunct}{\mcitedefaultseppunct}\relax
\EndOfBibitem
\bibitem[Kowalewski and Seeber(2022)Kowalewski, and Seeber]{nix}
Kowalewski,~M.; Seeber,~P. Sustainable packaging of quantum chemistry software
  with the Nix package manager. \emph{Int. J. Quant. Chem.} \textbf{2022},
  \emph{122}, e26872\relax
\mciteBstWouldAddEndPuncttrue
\mciteSetBstMidEndSepPunct{\mcitedefaultmidpunct}
{\mcitedefaultendpunct}{\mcitedefaultseppunct}\relax
\EndOfBibitem
\bibitem[Dunning(1989)]{Dunning1989-xc}
Dunning,~T.~H.,~Jr Gaussian basis sets for use in correlated molecular
  calculations. {I}. The atoms boron through neon and hydrogen. \emph{J. Chem.
  Phys.} \textbf{1989}, \emph{90}, 1007--1023\relax
\mciteBstWouldAddEndPuncttrue
\mciteSetBstMidEndSepPunct{\mcitedefaultmidpunct}
{\mcitedefaultendpunct}{\mcitedefaultseppunct}\relax
\EndOfBibitem
\bibitem[Fischer \latin{et~al.}(2024)Fischer, Syska, and
  Saalfrank]{Fischer2024-mk}
Fischer,~E.~W.; Syska,~J.~A.; Saalfrank,~P. A Quantum Chemistry Approach to
  Linear Vibro-Polaritonic Infrared Spectra with Perturbative Electron-Photon
  Correlation. \emph{J. Phys. Chem. Lett.} \textbf{2024}, 2262--2269\relax
\mciteBstWouldAddEndPuncttrue
\mciteSetBstMidEndSepPunct{\mcitedefaultmidpunct}
{\mcitedefaultendpunct}{\mcitedefaultseppunct}\relax
\EndOfBibitem
\bibitem[Fiechter and Richardson(2024)Fiechter, and
  Richardson]{Fiechter2024-gl}
Fiechter,~M.~R.; Richardson,~J.~O. Understanding the cavity Born-Oppenheimer
  approximation. \emph{J. Chem. Phys.} \textbf{2024}, \emph{160}\relax
\mciteBstWouldAddEndPuncttrue
\mciteSetBstMidEndSepPunct{\mcitedefaultmidpunct}
{\mcitedefaultendpunct}{\mcitedefaultseppunct}\relax
\EndOfBibitem
\bibitem[George \latin{et~al.}(2016)George, Chervy, Shalabney, Devaux, Hiura,
  Genet, and Ebbesen]{George2016-sy}
George,~J.; Chervy,~T.; Shalabney,~A.; Devaux,~E.; Hiura,~H.; Genet,~C.;
  Ebbesen,~T.~W. Multiple Rabi Splittings under Ultrastrong Vibrational
  Coupling. \emph{Phys. Rev. Lett.} \textbf{2016}, \emph{117}, 153601\relax
\mciteBstWouldAddEndPuncttrue
\mciteSetBstMidEndSepPunct{\mcitedefaultmidpunct}
{\mcitedefaultendpunct}{\mcitedefaultseppunct}\relax
\EndOfBibitem
\bibitem[Wright \latin{et~al.}(2023)Wright, Nelson, and
  Weichman]{Wright2023-xx}
Wright,~A.~D.; Nelson,~J.~C.; Weichman,~M.~L. A versatile platform for
  gas-phase molecular polaritonics. \emph{J. Chem. Phys.} \textbf{2023},
  \emph{159}, 164202\relax
\mciteBstWouldAddEndPuncttrue
\mciteSetBstMidEndSepPunct{\mcitedefaultmidpunct}
{\mcitedefaultendpunct}{\mcitedefaultseppunct}\relax
\EndOfBibitem
\end{mcitethebibliography}
\providecommand{\latin}[1]{#1}
\makeatletter
\providecommand{\doi}
  {\begingroup\let\do\@makeother\dospecials
  \catcode`\{=1 \catcode`\}=2 \doi@aux}
\providecommand{\doi@aux}[1]{\endgroup\texttt{#1}}
\makeatother
\providecommand*\mcitethebibliography{\thebibliography}
\csname @ifundefined\endcsname{endmcitethebibliography}
  {\let\endmcitethebibliography\endthebibliography}{}

\end{document}